\documentclass[11pt,fleqn]{book}
\usepackage[top=3cm,bottom=3cm,left=3cm,right=3cm,headsep=10pt,a4paper]{geometry} 

\usepackage{graphicx} 
\graphicspath{{img/}} 

\usepackage{lipsum} 

\usepackage{varioref}
\usepackage{forest}
\usepackage{listings}
\usepackage{verbatim}
\usepackage{longtable}

\usepackage{tikz} 

\usepackage[english]{babel} 

\usepackage{enumitem} 
\setlist{nolistsep} 

\usepackage{booktabs} 

\usepackage{xcolor} 
\definecolor{hesscolor}{RGB}{211, 27, 23}

\usepackage{avant} 
\usepackage{mathptmx} 

\usepackage{microtype} 
\usepackage[utf8]{inputenc} 
\usepackage[T1]{fontenc} 

\usepackage[citestyle=numeric,sorting=none,sortcites=true,autopunct=true,babel=hyphen,hyperref=true,abbreviate=false,backref=true,backend=biber]{biblatex}
\defbibheading{bibempty}{}

\usepackage{calc} 
\usepackage{makeidx} 
\makeindex 

\usepackage{titletoc} 

\contentsmargin{0cm} 

\titlecontents{part}[0cm]
{\addvspace{20pt}\centering\large\bfseries}
{}
{}
{}

\titlecontents{chapter}[1.25cm] 
{\addvspace{12pt}\large\sffamily\bfseries} 
{\color{hesscolor!60}\contentslabel[\Large\thecontentslabel]{1.25cm}\color{hesscolor}} 
{\color{hesscolor}}
{\color{hesscolor!60}\normalsize\;\titlerule*[.5pc]{.}\;\thecontentspage} 

\titlecontents{section}[1.25cm] 
{\addvspace{3pt}\sffamily\bfseries} 
{\contentslabel[\thecontentslabel]{1.25cm}} 
{}
{\hfill\color{black}\thecontentspage} 
[]

\titlecontents{subsection}[1.25cm] 
{\addvspace{1pt}\sffamily\small} 
{\contentslabel[\thecontentslabel]{1.25cm}} 
{}
{\ \titlerule*[.5pc]{.}\;\thecontentspage} 
[]

\titlecontents{figure}[0em]
{\addvspace{-5pt}\sffamily}
{\thecontentslabel\hspace*{1em}}
{}
{\ \titlerule*[.5pc]{.}\;\thecontentspage}
[]

\titlecontents{table}[0em]
{\addvspace{-5pt}\sffamily}
{\thecontentslabel\hspace*{1em}}
{}
{\ \titlerule*[.5pc]{.}\;\thecontentspage}
[]

\titlecontents{lchapter}[0em] 
{\addvspace{15pt}\large\sffamily\bfseries} 
{\color{hesscolor}\contentslabel[\Large\thecontentslabel]{1.25cm}\color{hesscolor}} 
{}
{\color{hesscolor}\normalsize\sffamily\bfseries\;\titlerule*[.5pc]{.}\;\thecontentspage} 

\titlecontents{lsection}[0em] 
{\sffamily\small} 
{\contentslabel[\thecontentslabel]{1.25cm}} 
{}
{}

\titlecontents{lsubsection}[.5em] 
{\normalfont\footnotesize\sffamily} 
{}
{}
{}

\usepackage{fancyhdr} 

\fancypagestyle{plain}{\fancyhead{}} 

\makeatletter
\renewcommand{\cleardoublepage}{
\clearpage\ifodd\c@page\else
\hbox{}
\vspace*{\fill}
\thispagestyle{empty}
\newpage
\fi}

\usepackage{amsmath,amsfonts,amssymb,amsthm} 

\newtheoremstyle{hesscolornumbox}
{0pt}
{0pt}
{\normalfont}
{}
{\small\bf\sffamily\color{hesscolor}}
{\;}
{0.25em}
{\small\sffamily\color{hesscolor}\thmname{#1}\nobreakspace\thmnumber{\@ifnotempty{#1}{}\@upn{#2}}
\thmnote{\nobreakspace\the\thm@notefont\sffamily\bfseries\color{black}---\nobreakspace#3.}} 

\newtheoremstyle{blacknumex}
{5pt}
{5pt}
{\normalfont}
{} 
{\small\bf\sffamily}
{\;}
{0.25em}
{\small\sffamily{\tiny\ensuremath{\blacksquare}}\nobreakspace\thmname{#1}\nobreakspace\thmnumber{\@ifnotempty{#1}{}\@upn{#2}}
\thmnote{\nobreakspace\the\thm@notefont\sffamily\bfseries---\nobreakspace#3.}}

\newtheoremstyle{blacknumbox} 
{0pt}
{0pt}
{\normalfont}
{}
{\small\bf\sffamily}
{\;}
{0.25em}
{\small\sffamily\thmname{#1}\nobreakspace\thmnumber{\@ifnotempty{#1}{}\@upn{#2}}
\thmnote{\nobreakspace\the\thm@notefont\sffamily\bfseries---\nobreakspace#3.}}

\newtheoremstyle{hesscolornum}
{5pt}
{5pt}
{\normalfont}
{}
{\small\bf\sffamily\color{hesscolor}}
{\;}
{0.25em}
{\small\sffamily\color{hesscolor}\thmname{#1}\nobreakspace\thmnumber{\@ifnotempty{#1}{}\@upn{#2}}
\thmnote{\nobreakspace\the\thm@notefont\sffamily\bfseries\color{black}---\nobreakspace#3.}} 
\makeatother

\newcounter{dummy}
\numberwithin{dummy}{section}
\theoremstyle{hesscolornumbox}
\newtheorem{theoremeT}[dummy]{Theorem}

\newtheorem{exerciseT}{Exercise}[chapter]
\theoremstyle{blacknumex}
\newtheorem{exampleT}{Example}[chapter]
\theoremstyle{blacknumbox}

\newtheorem{definitionT}{Definition}[section]
\newtheorem{corollaryT}[dummy]{Corollary}
\theoremstyle{hesscolornum}

\RequirePackage[framemethod=default]{mdframed} 

\newmdenv[skipabove=7pt,
skipbelow=7pt,
backgroundcolor=black!5,
linecolor=hesscolor,
innerleftmargin=5pt,
innerrightmargin=5pt,
innertopmargin=5pt,
leftmargin=0cm,
rightmargin=0cm,
innerbottommargin=5pt]{tBox}

\newmdenv[skipabove=7pt,
skipbelow=7pt,
rightline=false,
leftline=true,
topline=false,
bottomline=false,
backgroundcolor=hesscolor!10,
linecolor=hesscolor,
innerleftmargin=5pt,
innerrightmargin=5pt,
innertopmargin=5pt,
innerbottommargin=5pt,
leftmargin=0cm,
rightmargin=0cm,
linewidth=4pt]{eBox}

\newmdenv[skipabove=7pt,
skipbelow=7pt,
rightline=false,
leftline=true,
topline=false,
bottomline=false,
linecolor=hesscolor,
innerleftmargin=5pt,
innerrightmargin=5pt,
innertopmargin=0pt,
leftmargin=0cm,
rightmargin=0cm,
linewidth=4pt,
innerbottommargin=0pt]{dBox}

\newmdenv[skipabove=7pt,
skipbelow=7pt,
rightline=false,
leftline=true,
topline=false,
bottomline=false,
linecolor=gray,
backgroundcolor=black!5,
innerleftmargin=5pt,
innerrightmargin=5pt,
innertopmargin=5pt,
leftmargin=0cm,
rightmargin=0cm,
linewidth=4pt,
innerbottommargin=5pt]{cBox}

\makeatletter
\renewcommand{\@seccntformat}[1]{\llap{\textcolor{hesscolor}{\csname the#1\endcsname}\hspace{1em}}}
\renewcommand{\section}{\@startsection{section}{1}{\z@}
{-4ex \@plus -1ex \@minus -.4ex}
{1ex \@plus.2ex }
{\normalfont\large\sffamily\bfseries}}
\renewcommand{\subsection}{\@startsection {subsection}{2}{\z@}
{-3ex \@plus -0.1ex \@minus -.4ex}
{0.5ex \@plus.2ex }
{\normalfont\sffamily\bfseries}}
\renewcommand{\subsubsection}{\@startsection {subsubsection}{3}{\z@}
{-2ex \@plus -0.1ex \@minus -.2ex}
{.2ex \@plus.2ex }
{\normalfont\small\sffamily\bfseries}}
\renewcommand\paragraph{\@startsection{paragraph}{4}{\z@}
{-2ex \@plus-.2ex \@minus .2ex}
{.1ex}
{\normalfont\small\sffamily\bfseries}}

\newcommand{\@mypartnumtocformat}[2]{%
\setlength\fboxsep{0pt}%
\noindent\colorbox{hesscolor!20}{\strut\parbox[c][.7cm]{\ecart}{\color{hesscolor!70}\Large\sffamily\bfseries\centering#1}}\hskip\esp\colorbox{hesscolor!40}{\strut\parbox[c][.7cm]{\linewidth-\ecart-\esp}{\Large\sffamily\centering#2}}}%
\newcommand{\@myparttocformat}[1]{%
\setlength\fboxsep{0pt}%
\noindent\colorbox{hesscolor!40}{\strut\parbox[c][.7cm]{\linewidth}{\Large\sffamily\centering#1}}}%
\newlength\esp
\newlength\ecart
\def\@part[#1]#2{%
\ifnum \c@secnumdepth >-2\relax%
\refstepcounter{part}%
\addcontentsline{toc}{part}{\texorpdfstring{\protect\@mypartnumtocformat{\thepart}{#1}}{\partname~\thepart\ ---\ #1}}
\else%
\addcontentsline{toc}{part}{\texorpdfstring{\protect\@myparttocformat{#1}}{#1}}%
\fi%
\startcontents%
\markboth{}{}%
{\thispagestyle{empty}%
\begin{tikzpicture}[remember picture,overlay]%
\node at (current page.north west){\begin{tikzpicture}[remember picture,overlay]%
\fill[hesscolor!20](0cm,0cm) rectangle (\paperwidth,-\paperheight);
\node[anchor=north] at (4cm,-3.25cm){\color{hesscolor!40}\fontsize{220}{100}\sffamily\bfseries\@Roman\c@part};
\node[anchor=south east] at (\paperwidth-1cm,-\paperheight+1cm){\parbox[t][][t]{8.5cm}{
\printcontents{l}{0}{\setcounter{tocdepth}{1}}%
}};
\node[anchor=north east] at (\paperwidth-1.5cm,-3.25cm){\parbox[t][][t]{15cm}{\strut\raggedleft\color{white}\fontsize{30}{30}\sffamily\bfseries#2}};
\end{tikzpicture}};
\end{tikzpicture}}%
\@endpart}
\def\@spart#1{%
\startcontents%
\phantomsection
{\thispagestyle{empty}%
\begin{tikzpicture}[remember picture,overlay]%
\node at (current page.north west){\begin{tikzpicture}[remember picture,overlay]%
\fill[hesscolor!20](0cm,0cm) rectangle (\paperwidth,-\paperheight);
\node[anchor=north east] at (\paperwidth-1.5cm,-3.25cm){\parbox[t][][t]{15cm}{\strut\raggedleft\color{white}\fontsize{30}{30}\sffamily\bfseries#1}};
\end{tikzpicture}};
\end{tikzpicture}}
\addcontentsline{toc}{part}{\texorpdfstring{%
\setlength\fboxsep{0pt}%
\noindent\protect\colorbox{hesscolor!40}{\strut\protect\parbox[c][.7cm]{\linewidth}{\Large\sffamily\protect\centering #1\quad\mbox{}}}}{#1}}%
\@endpart}
\def\@endpart{\vfil\newpage
\if@twoside
\if@openright
\null
\thispagestyle{empty}%
\newpage
\fi
\fi
\if@tempswa
\twocolumn
\fi}

\newif\ifusechapterimage
\usechapterimagetrue
\newcommand{\thechapterimage}{}%
\newcommand{\chapterimage}[1]{\ifusechapterimage\renewcommand{\thechapterimage}{#1}\fi}%
\def\@makechapterhead#1{%
{\parindent \z@ \raggedright \normalfont
\ifnum \c@secnumdepth >\m@ne
\if@mainmatter
\begin{tikzpicture}[remember picture,overlay]
\node at (current page.north west)
{\begin{tikzpicture}[remember picture,overlay]
\node[anchor=north west,inner sep=0pt] at (0,0) {\ifusechapterimage\includegraphics[width=\paperwidth]{\thechapterimage}\fi};
\draw[anchor=west] (\Gm@lmargin,-9cm) node [line width=2pt,rounded corners=15pt,draw=hesscolor,fill=white,fill opacity=0.5,inner sep=15pt]{\strut\makebox[22cm]{}};
\draw[anchor=west] (\Gm@lmargin+.3cm,-9cm) node {\huge\sffamily\bfseries\color{black}\thechapter. #1\strut};
\end{tikzpicture}};
\end{tikzpicture}
\else
\begin{tikzpicture}[remember picture,overlay]
\node at (current page.north west)
{\begin{tikzpicture}[remember picture,overlay]
\node[anchor=north west,inner sep=0pt] at (0,0) {\ifusechapterimage\includegraphics[width=\paperwidth]{\thechapterimage}\fi};
\draw[anchor=west] (\Gm@lmargin,-9cm) node [line width=2pt,rounded corners=15pt,draw=hesscolor,fill=white,fill opacity=0.5,inner sep=15pt]{\strut\makebox[22cm]{}};
\draw[anchor=west] (\Gm@lmargin+.3cm,-9cm) node {\huge\sffamily\bfseries\color{black}#1\strut};
\end{tikzpicture}};
\end{tikzpicture}
\fi\fi\par\vspace*{270\p@}}}

\def\@makeschapterhead#1{%
\begin{tikzpicture}[remember picture,overlay]
\node at (current page.north west)
{\begin{tikzpicture}[remember picture,overlay]
\node[anchor=north west,inner sep=0pt] at (0,0) {\ifusechapterimage\includegraphics[width=\paperwidth]{\thechapterimage}\fi};
\draw[anchor=west] (\Gm@lmargin,-9cm) node [line width=2pt,rounded corners=15pt,draw=hesscolor,fill=white,fill opacity=0.5,inner sep=15pt]{\strut\makebox[22cm]{}};
\draw[anchor=west] (\Gm@lmargin+.3cm,-9cm) node {\huge\sffamily\bfseries\color{black}#1\strut};
\end{tikzpicture}};
\end{tikzpicture}
\par\vspace*{270\p@}}
\makeatother

\usepackage{hyperref}
\hypersetup{hidelinks,backref=true,pagebackref=true,hyperindex=true,colorlinks=false,breaklinks=true,urlcolor= hesscolor,bookmarks=true,bookmarksopen=false,pdftitle={Title},pdfauthor={Author}}
\usepackage{bookmark}
\bookmarksetup{
open,
numbered,
addtohook={%
\ifnum\bookmarkget{level}=0 
\bookmarksetup{bold}%
\fi
\ifnum\bookmarkget{level}=-1 
\bookmarksetup{color=hesscolor,bold}%
\fi
}
}

\usepackage{acronym}
\usepackage{subcaption}
\usepackage{xspace}
\usepackage{authblk}

\newcommand{\crab}{\mbox{Crab~nebula}\xspace}
\newcommand{\pks}{\mbox{PKS~2155$-$304}\xspace}
\newcommand{\msh}{\mbox{MSH~15$-$52}\xspace}
\newcommand{\rxj}{\mbox{RX~J1713.7$-$3946}\xspace}

\newcommand{\hess}{H.E.S.S.\xspace}
\newcommand{\hessOne}{H.E.S.S.~I\xspace}

\newcommand{\urlFits}{\url{http://fits.gsfc.nasa.gov/}\xspace}
\newcommand{\urlOgip}{\url{https://heasarc.gsfc.nasa.gov/docs/heasarc/ofwg/ofwg_intro.html}\xspace}
\newcommand{\urlGadfGithub}{\url{https://github.com/open-gamma-ray-astro/gamma-astro-data-formats}\xspace}
\newcommand{\urlGadfDocs}{\url{https://gamma-astro-data-formats.readthedocs.io/}\xspace}
\newcommand{\urlHESS}{\url{https://www.mpi-hd.mpg.de/hfm/HESS/}\xspace}
\newcommand{\urlRelease}{\url{https://www.mpi-hd.mpg.de/hfm/HESS/pages/dl3-dr1/}\xspace}

\newcommand{\urlGammapy}{\url{http://gammapy.org/}}
\newcommand{\urlCtools}{\url{http://cta.irap.omp.eu/ctools/}}

\newcommand{\hapVersionDate}{May 6, 2018\xspace}
\newcommand{\astropyVersion}{3.1\xspace}
\newcommand{\gammapyVersion}{0.8\xspace}
\newcommand{\gadfVersion}{0.2\xspace}
\newcommand{\dstVersion}{12-03}

\newcommand{\crabIrfDescr}{zenith angle 49~deg, FOV offset 0.5~deg, safe energy
threshold 0.69~TeV}

\newcommand{\sourceImageDescr}{counts image (left) and counts spectrum (right)
for a circular on region (total in red, background estimate using ring method in
blue).}

\newcommand{\dataRunsTotal}{105}
\newcommand{\dataTimeSources}{27.9}
\newcommand{\dataRunsSources}{60}
\newcommand{\dataTimeOff}{20.7}
\newcommand{\dataRunsOff}{45}
\newcommand{\dataTimeCrab}{1.9}
\newcommand{\dataRunsCrab}{4}
\newcommand{\dataTimePksSteady}{2.8}
\newcommand{\dataRunsPksSteady}{6}
\newcommand{\dataTimePksFlare}{7.0}
\newcommand{\dataRunsPksFlare}{15}
\newcommand{\dataTimeMsh}{9.1}
\newcommand{\dataRunsMsh}{20}
\newcommand{\dataTimeRxj}{7.0}
\newcommand{\dataRunsRxj}{15}
\newcommand{\dataEventsCountTotal}{1,046,156} 
\newcommand{\dataEventsCountMean}{9,963} 
\newcommand{\dataObsIndexCols}{31}
\newcommand{\dataHduIndexCols}{7}

\newcommand{\dataSizeEvents}{360.1}
\newcommand{\dataSizeGti}{5.6}
\newcommand{\dataSizeAeff}{11.2}
\newcommand{\dataSizeEdisp}{368.4}
\newcommand{\dataSizePsf}{115.3}
\newcommand{\dataSizeFitsDataAll}{42.8}

\begin{document}

\addtolength{\intextsep}{+0.3cm}

\frontmatter
\begingroup
\thispagestyle{empty}
\begin{tikzpicture}[remember picture,overlay]
\node at (current page.north west)
{\begin{tikzpicture}[remember picture,overlay]
\node[anchor=north west,inner sep=0pt] at (0,0) {\includegraphics[width=\paperwidth]{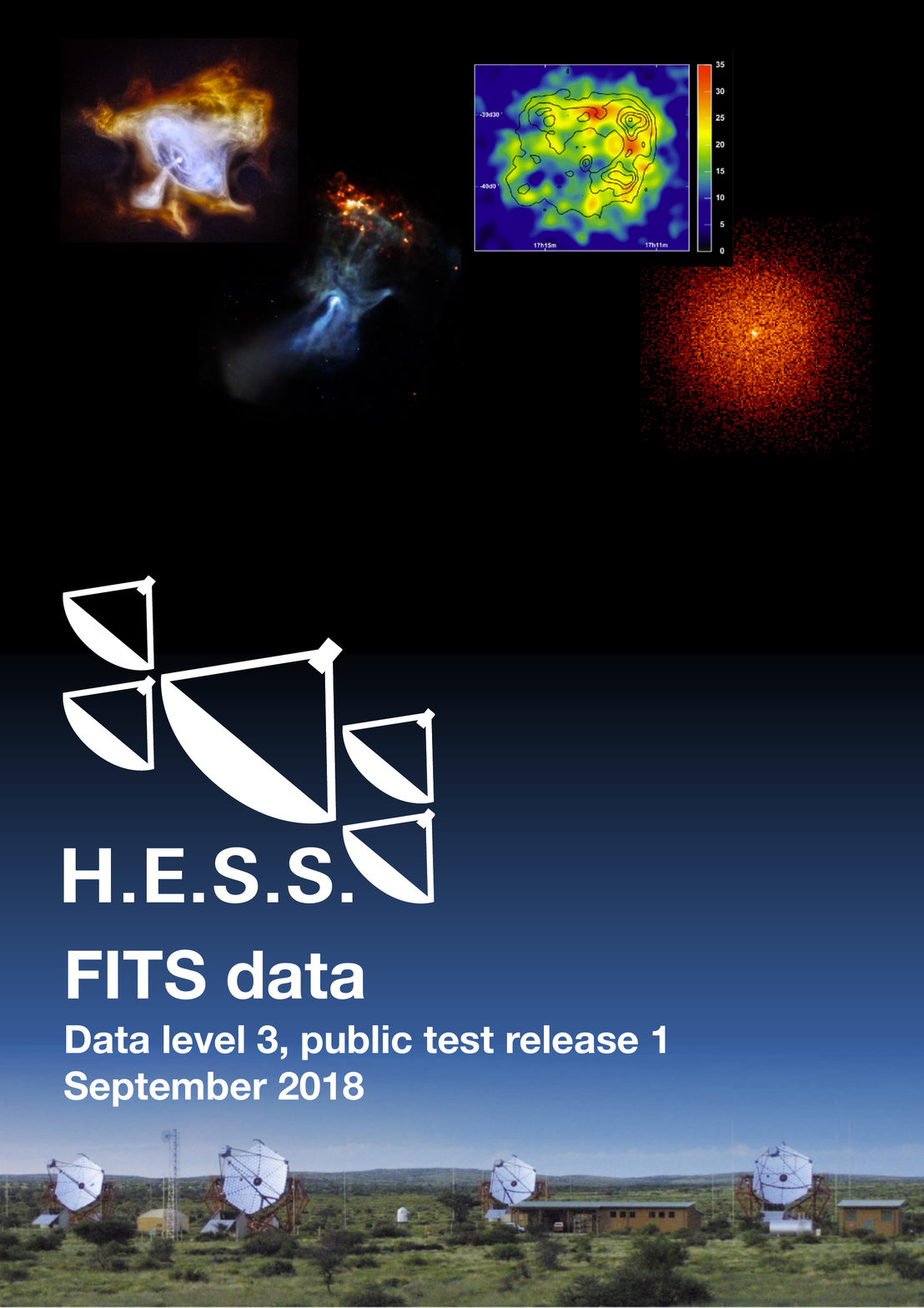}}; 
\end{tikzpicture}};
\end{tikzpicture}
\vfill
\endgroup


\leavevmode\thispagestyle{empty}\newpage
\thispagestyle{empty}

\title{}
\date{}

\author[$\,$]{H.E.S.S.~Collaboration}
\author[1]{H.~Abdalla} 
\author[2]{A.~Abramowski} 
\author[3,4,5]{F.~Aharonian} 
\author[3]{F.~Ait~Benkhali} 
\author[6]{E.O.~Ang\"uner} 
\author[7]{M.~Arakawa} 
\author[1]{C.~Arcaro} 
\author[8]{C.~Armand} 
\author[9]{M.~Arrieta} 
% ...

\affil[1]{Centre for Space Research, North-West University, Potchefstroom 2520, South Africa} 
\affil[2]{Universit\"at Hamburg, Institut f\"ur Experimentalphysik, Luruper Chaussee 149, D 22761 Hamburg, Germany} 
\affil[3]{Max-Planck-Institut f\"ur Kernphysik, P.O. Box 103980, D 69029 Heidelberg, Germany} 
\affil[4]{Dublin Institute for Advanced Studies, 31 Fitzwilliam Place, Dublin 2, Ireland} 
\affil[5]{National Academy of Sciences of the Republic of Armenia,  Marshall Baghramian Avenue, 24, 0019 Yerevan, Republic of Armenia} 
\affil[6]{Aix Marseille Universit\'e, CNRS/IN2P3, CPPM, Marseille, France} 
\affil[7]{Department of Physics, Rikkyo University, 3-34-1 Nishi-Ikebukuro, Toshima-ku, Tokyo 171-8501, Japan} 
\affil[8]{Laboratoire d'Annecy de Physique des Particules, Univ. Grenoble Alpes, Univ. Savoie Mont Blanc, CNRS, LAPP, 74000 Annecy, France} 
\affil[9]{LUTH, Observatoire de Paris, PSL Research University, CNRS, Universit\'e Paris Diderot, 5 Place Jules Janssen, 92190 Meudon, France} 
% ...

{\let\newpage\relax\pagestyle{empty}\maketitle}
\thispagestyle{empty}

\noindent \textbf{Date: September 18, 2018}

\vspace{1 cm}

\noindent \textbf{Abstract}

\vspace{0.5 cm}

The High Energy Stereoscopic System (\hess) is an array of ground-based imaging
atmospheric Cherenkov telescopes in Namibia. For the first time, the \hess
collaboration is releasing a small dataset of event lists and instrument
response information. This is a test data release, with the motivation to
support the ongoing efforts to define open high-level data models and associated
formats, as well as open-source science tools for gamma-ray astronomy. The data
are in FITS format. Open-source science tools that support this format exist
already.

The release data consists of \dataTimeSources~hours in total of observations of
the \crab, \pks, \msh and \rxj taken with the \hess~1 array. Most data are from
2004, the \pks data are from 2006 and 2008. In addition, \dataTimeOff~hours of
off observations of empty fields of view are included. The targets and
observations were chosen to be suitable for common analysis use cases, including
point-like and extended sources for spectral and morphology measurements, as
well as a variable source (\pks) and the off dataset for background studies. The
total size of the files in this data release is \dataSizeFitsDataAll~MB.

This is a very small subset of the thousands of hours of \hess~1 observations
taken since 2004. The quality of this dataset, and measurements derived from
this data, does not reflect the state of the art for \hess\ publications, e.g.
the event reconstruction and gamma-hadron separation method used here is a very
basic one.

\vspace{1 cm}

\noindent Webpage: \urlRelease

\vspace{1 cm}

\noindent Questions or comments: \href{mailto:contact.hess@hess-experiment.eu}{contact.hess@hess-experiment.eu}.

\vspace{1 cm}

\noindent This data release was prepared by the H.E.S.S. FITS data task group.
Members (current and former) include: Christoph Deil, Lars Mohrmann, Johannes
King, Catherine Boisson, Axel Donath, Julien Lefaucheur, Bruno Khélifi, Léa
Jouvin, Régis Terrier, Alexander Ziegler, Domenico Tiziani, Christopher Sobel,
Karl Kosack, Michael Mayer and Anneli Schulz.

\leavevmode\thispagestyle{empty}\newpage\leavevmode\thispagestyle{empty}\newpage
\newpage
\thispagestyle{empty}

This data is released under the terms of use stated in the \verb=README.txt=
file, which is included here verbatim:

\begin{verbatim}
H.E.S.S. DL3 public test data release 1 (HESS DL3 DR1)
------------------------------------------------------

H.E.S.S. collaboration, 2018

Webpage: https://www.mpi-hd.mpg.de/hfm/HESS/pages/dl3-dr1/

The data and documentation is publicly released by the H.E.S.S. collaboration
as a contribution to the ongoing efforts to define a common open format for
data level 3 of imaging atmospheric Cherenkov telescopes (IACTs) and IACT
open-source science tool development, to enlarge the community involved in
IACT data analysis.

No scientific publications may be derived from the data. Using the data
for new claims about the astrophysical sources is not permitted.

When using this data, please include the following attribution:

    This work made use of data from the H.E.S.S. DL3 public test
    data release 1 (HESS DL3 DR1, H.E.S.S. collaboration, 2018).

Alternatively, use the following shorter version, e.g. for presentations:

    HESS DL3 DR1, H.E.S.S. collaboration

These terms of use must be included in all copies in full or part of the data.

For information on context, aims, use and contacts, as well as a description
of the dataset, see the hess_dl3_dr1.pdf document.
\end{verbatim}



\chapterimage{layout/chapter_head3_small}
\pagestyle{empty} 

\setcounter{tocdepth}{1}
\tableofcontents


\pagestyle{fancy} 

\mainmatter

\chapterimage{layout/chapter_head8_small}
\chapter{Introduction}
\label{sec:intro}

\section{\hess}
\label{sec:ds:hess}

The High Energy Stereoscopic System (\hess)\footnote{\urlHESS} is an array of
imaging air Cherenkov telescopes (IACTs) situated in the Khomas Highland,
Namibia, at 1800~meter above sea level. Since 2004, four telescopes (\hess~Phase
I) with mirror surfaces of $\sim 100$~m$^2$ have been detecting air
showers produced by $\gamma$ rays in the 100~GeV to 100~TeV energy band. This
array forms a square of 120~m side length. It has a field of view of 5~deg in
diameter, a spatial resolution of $\sim 0.1$~deg and an energy resolution of
$\sim 15\%$ \cite{Aharonian:2006f}. In September 2012, a fifth telescope placed
in the middle of the original square was inaugurated, initiating \hess Phase
II. It has a mirror surface of $\sim 600$~m$^2$ and lowers the energy
threshold of \hess to tens of GeV. The data in this release were taken mostly in
2004 (some in 2005-2008), all with the four \hess~1 telescopes.

\section{Context}
\label{sec:context}

Ground-based gamma-ray astronomy is a relatively new window on the cosmos. The
existing ground-based IACTs like e.g.
\hess, MAGIC and VERITAS, have been operating independently for the past decade,
using proprietary data formats and codes. The Cherenkov Telescope Array (CTA),
the next IACT instrument, will probe the non-thermal universe above 20 GeV up to
a few 100 TeV with an unmatched sensitivity and angular resolution compared to
the current IACT experiments. CTA will be the first ground-based gamma-ray
telescope array operated as an open observatory  with public observer access.
This implies fundamentally different requirements for the data formats and
software tools and a challenge on their implementation to make very high energy
(VHE) gamma-ray astronomy as accessible as any other waveband.

The Flexible Image Transport System (FITS) has been used by
astronomers as a data interchange and archiving format for decades
(\cite{Pence:2010}, \urlFits).  Space missions in X-ray or high-energy astronomy
also store the list of recorded events, containing information like their
arrival direction, time and energy, in FITS file format\footnote{\urlOgip}. This
is not yet the case within the VHE astronomy community, particularly among the
international collaborations operating ground-based IACTs, due to the different
culture regarding data and software distribution in the particle physics community
compared to the astrophysics one.

With that in mind and following on the work in CTA to provide DL3 event lists
and IRF information in FITS format (\cite{2016arXiv161001884D}), \hess
Collaboration members have written exporters for DL3 data from
different reconstruction chains to FITS format for internal use. The current DL3
data model and format definition needs to be tested on observations, to judge whether
it properly specifies all data necessary for high-level science analysis. Such an
effort is thus a valuable test bench and input for discussion in view of CTA.

Agreeing on a common data format for files greatly simplifies mid-level (event energies,
positions) and high-level (source position, morphology, spectrum) checks between
the different chains, algorithms and open-source tools. This will also
ease interoperability with other codes (e.g.~to check results, combine results
in one plot, \ldots). Currently two open-source science tools packages are being
designed for current IACT and CTA data analysis, Gammapy
\cite{2015arXiv150907408D, 2017arXiv170901751D} and ctools
\cite{2016AnA...593A...1K}. \emph{Gammapy} is an in-development
Astropy-affiliated  package, which is mainly written in Python, and
\emph{ctools} is based on the GammaLib analysis framework, which is mainly
written in C++. Both can read as input the format used for this \hess data
release.

The format specifications currently being developed can form the basis for
prototyping for data producers (i.e.\ existing IACTs and simulated CTA data)
and consumers (i.e.\ science tool codes). They are made accessible on
Github\footnote{\urlGadfGithub}, so that they are visible by all, not just CTA
members. Such open specifications also provide a basis to discuss the proposed
DL3 model and format with members of other IACT experiments, aiming towards a
common standard how to archive, diffuse and consume DL3 data.

\section{Aims}
\label{sec:aims}

The goal of this high-level data release (event lists and IRFs for high-level
science analysis) is to have real VHE data publicly available for
software and analysis method testing, and not to release data for science
analysis.

A small set of data taken on TeV sources is made available. This dataset is
prepared to allow a larger community to get their hands on VHE data, and to
give feedback on the format and open-source tools. This will allow to explore
requirements for analysis software.

VHE astronomers will have the opportunity to explore their familiar data with
the open-source science tools currently designed for analysis of astronomical
gamma-ray data, and also with standard tools used in the analysis of data in
other wavelengths (e.g.~ftools, xspec,~\ldots). Astronomers, who have not yet
worked in the field of VHE astronomy, will get a first view on the details of
VHE data analysis while dealing with familiar format and tools.

This will also be beneficial for CTA, which can profit from the experience gained
by this release.

\chapterimage{layout/chapter_head7_small}
\chapter{Dataset}
\label{sec:ds}

\section{Overview}
\label{sec:ds:sources}

The data released consist of \dataTimeSources~hours of observations of the
\crab, \pks, \msh and \rxj taken with the \hess~1 array. Most data are from 2004,
the \pks data are from 2006 and 2008. In addition, \dataTimeOff~hours of off
observations of empty fields of view are included. This is a very small subset
of the thousands of hours of \hess~1 observations taken since 2004. The targets and
observations were chosen to be suitable for common analysis use cases, including
point-like and extended sources for spectral and morphology measurements, as well
as a variable source (\pks) and the off dataset for background studies.

The data consist of so-called ``runs'', which are observations of usually
28~min duration (sometimes less) on a fixed RA/DEC position in the sky,
identified by a unique observation identifier (\texttt{OBS\_ID}). This release
contains \dataRunsTotal~runs in total: \dataRunsSources~runs observing a gamma-ray source and \dataRunsOff\ off runs. There are \dataEventsCountTotal\ events (on average
\dataEventsCountMean\ per run), most of which are cosmic-ray air-shower
background events.

A summary of all sources and observations is given in Tables~\ref{tab:sources},
\ref{tab:obs_summary} and Figure~\ref{fig:ds:obs}. The complete run list
(grouped by source) is given in Table~\vref{tab:obs_list} in
Appendix~\ref{sec:appendix:obs_list}. Table~\ref{tab:events} contains a summary
of the available event statistics for each source, i.e. an estimate of the
number of excess gamma-ray events and the background level.

The following sections give some information on each source, as well as a counts
image and spectrum (stacked for all observations) illustrating the
spatial and energy distribution of the events.

\begin{table}[tb]
\centering
\begin{tabular}{ lrrlll }
Source name &       RA &       DEC & Type  & Size        & H.E.S.S. publications \\\midrule
\crab       &    83.63 &     22.01 & PWN   & Point-like  & \cite{Aharonian:2006f}, \cite{2017arXiv170704196H} \\
\pks        &   329.72 &  $-$30.23 & AGN   & Point-like  & \cite{2009ApJ...696L.150A}, \cite{2009AaA...502..749A}, \cite{2012AaA...539A.149H} \\
\msh        &   228.53 &  $-$59.16 & PWN   & Small       & \cite{Aharonian:2005c}, \cite{2017arXiv170901422T} \\
\rxj        &   258.35 &     39.77 & SNR   & Large       & \cite{Aharonian:2006e}, \cite{Aharonian:2007c}, \cite{2016arXiv160908671H} \\
\end{tabular}
\caption[Sources included in this release]{
Sources included in this release. The H.E.S.S. publications are only given for
reference, the datasets from these publications do not match the ones released
here. The positions given here are from SIMBAD for the \crab pulsar and the AGN for \pks,
the position of \msh is the best-fit position from  \cite{Aharonian:2005c}; for \rxj from \cite{2016arXiv160908671H}.
}
\label{tab:sources}
\end{table}

\begin{table}
\centering
\begin{tabular}{lrrc}
Source Name & $\mathrm N_{\mathrm{runs}}$ & Time (h) & Dates \\
\midrule
Crab                   &          4 &        1.9 & 2004-12-04 - 2004-12-08 \\
PKS 2155-304 (flare)   &         15 &        7.0 & 2006-07-29 - 2006-07-30 \\
PKS 2155-304 (steady)  &          6 &        2.8 & 2008-08-27 - 2008-08-28 \\
MSH 15-52              &         20 &        9.1 & 2004-03-26 - 2004-04-19 \\
RX J1713.7-3946        &         15 &        7.0 & 2004-04-17 - 2004-05-21 \\
Off data               &         45 &       20.7 & 2004-04-14 - 2005-11-20 \\

\end{tabular}
\caption[Observation dataset summary]{
Observation dataset summary. $\mathrm N_{\mathrm{runs}}$ is the number of
observations. The table gives the observation time in hours and the range of
dates when the observations took place. Information on the available event
statistics in these datasets is given in Table~\ref{tab:events}. A full list of
observations is given in Table~\ref{tab:obs_list}.
}
\label{tab:obs_summary}
\end{table}

\begin{table}[tb]
\centering
\begin{tabular}{ lrrrrrrrr }
\hline\hline
Source name &
$\theta$ &
$N^{all}$ &
$N^{on}$ &
$N^{on}_\gamma$ &
$N^{on}_{bkg}$ &
$S$ \\
\hline
Crab                   &      0.3 &     30,129 &    1,549 &  1,084.0 &    465.0 &     33.0 \\
PKS 2155-304 (flare)   &      0.3 &    141,715 &   24,164 & 21,549.9 &  2,614.1 &    197.0 \\
PKS 2155-304 (steady)  &      0.3 &     36,888 &    1,115 &    380.8 &    734.2 &     11.4 \\
MSH 15-52              &      0.3 &    227,830 &    5,963 &  1,459.5 &  4,503.5 &     18.2 \\
RX J1713.7-3946        &      0.5 &    226,264 &   16,696 &  3,652.1 & 13,043.9 &     24.2 \\

\hline
\end{tabular}
\caption[Event summary statistics information for each dataset] {
Event summary statistics information for each data set. (See
Table~\vref{tab:obs_summary} for the definition of each data set, and
Table~\vref{tab:sources} for the positions used for this measurement.) The
number of events in the total data set for each source is given as $N^{all}$ (no
energy or field of view offset cut). Using circular aperture photometry and a
simple ring background estimate, a source gamma-ray excess~$N^{on}_{\gamma} =
N^{on} - N^{on}_{bkg}$ and significance~$S$ estimate using equation~(17) from
Li~\&~Ma \cite{1983ApJ...272..317L} was obtained. The on-region size radius
$\theta$ (in deg) is given in the table. The background ring was chosen with
radius 0.4 -- 0.7~deg (0.6 -- 0.9~deg for \rxj), no acceptance correction was
applied.
}
\label{tab:events}
\end{table}

\begin{figure}[tb]
\centering
\includegraphics[width=\linewidth]{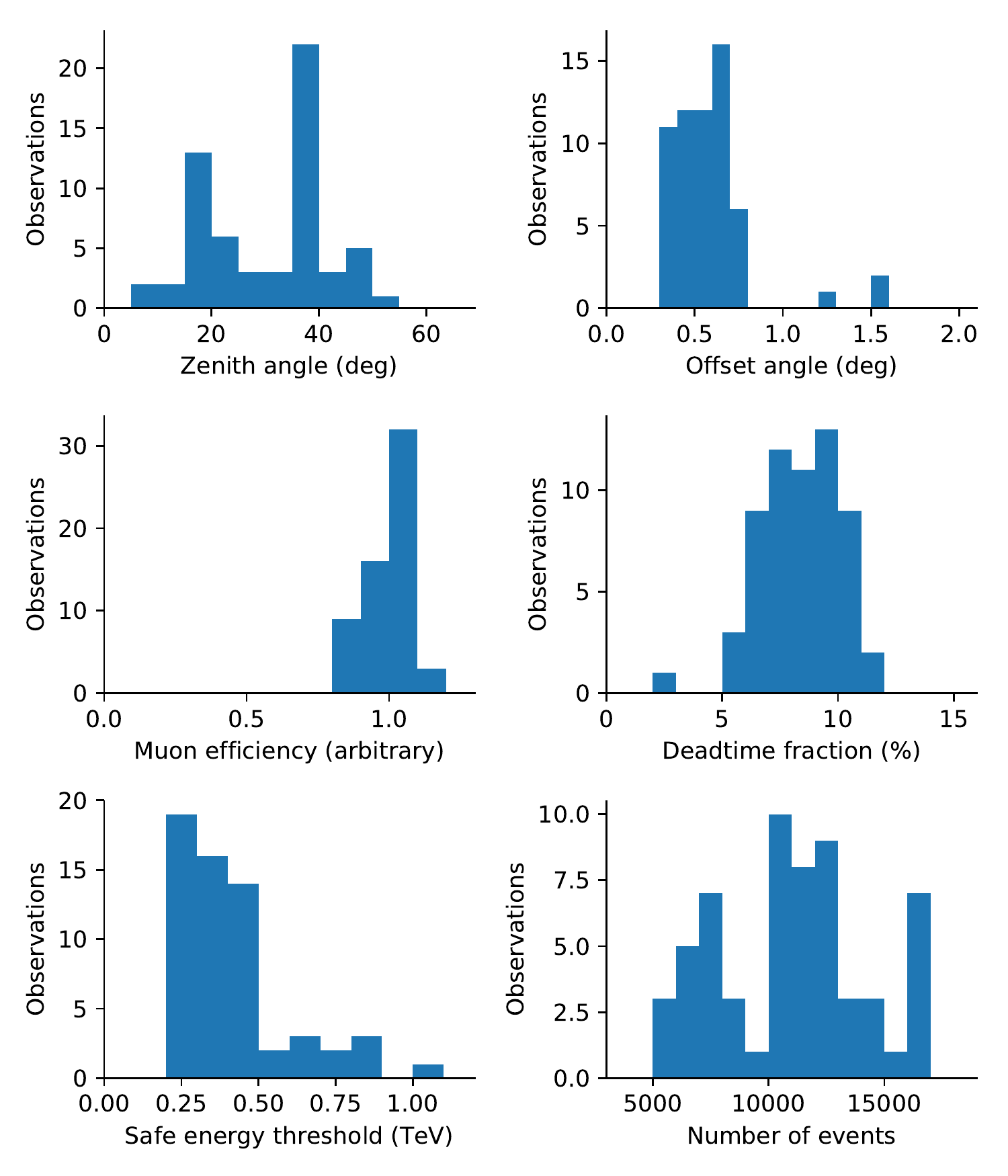}
\caption[Observation summary]{
Summary of parameters for the \dataRunsSources\ observation runs in this data
release. The parameters for the \dataRunsOff~off runs are not shown. Offset
angle is the sky separation between the pointing position and the target. Muon
efficiency is explained in Section~\ref{sec:notes:prod:irfs}, deadtime fraction
in Section~\ref{sec:df:events} and safe energy threshold in
Section~\ref{sec:notes:ana:cuts}. The number of events is for all energies and
the whole field of view.
}
\label{fig:ds:obs}
\end{figure}

\clearpage
\section{\crab}
\label{sec:ds:crab}

The \crab was the first VHE gamma-ray source detected and is one of the
brightest in the VHE sky. Gamma-ray emission has been detected from the pulsar
(dominating at GeV energies) and pulsar wind nebula (dominating at TeV
energies). Variability was detected at GeV energies \cite{Aliu:2008}. Recently,
the extension of the TeV nebula was measured \cite{2017arXiv170704196H}.

\begin{figure}[h] \centering
\begin{subfigure}{.44\textwidth}
 \includegraphics[height=6cm]{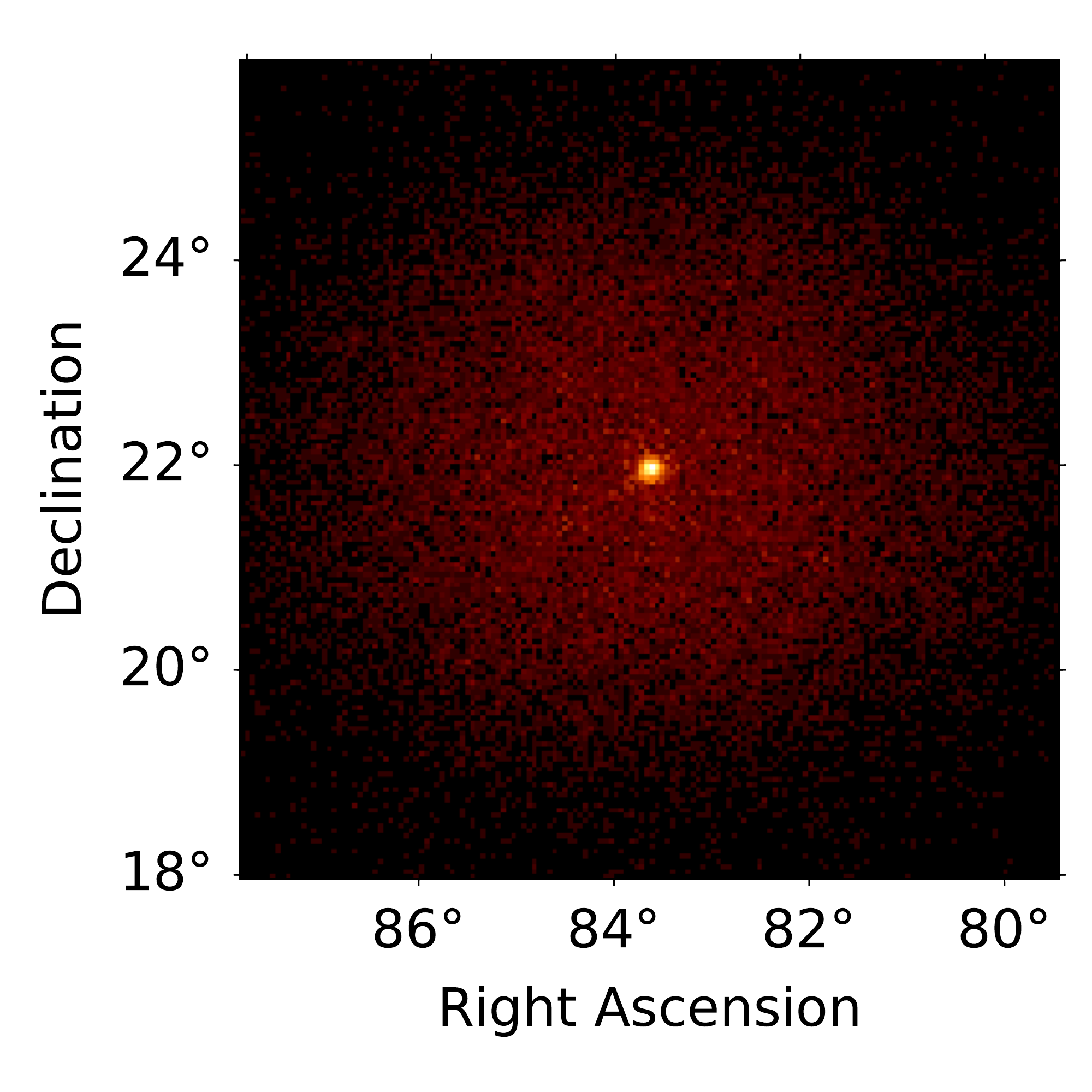}
\end{subfigure}
\begin{subfigure}{0.55\textwidth}
  \includegraphics[height=6cm]{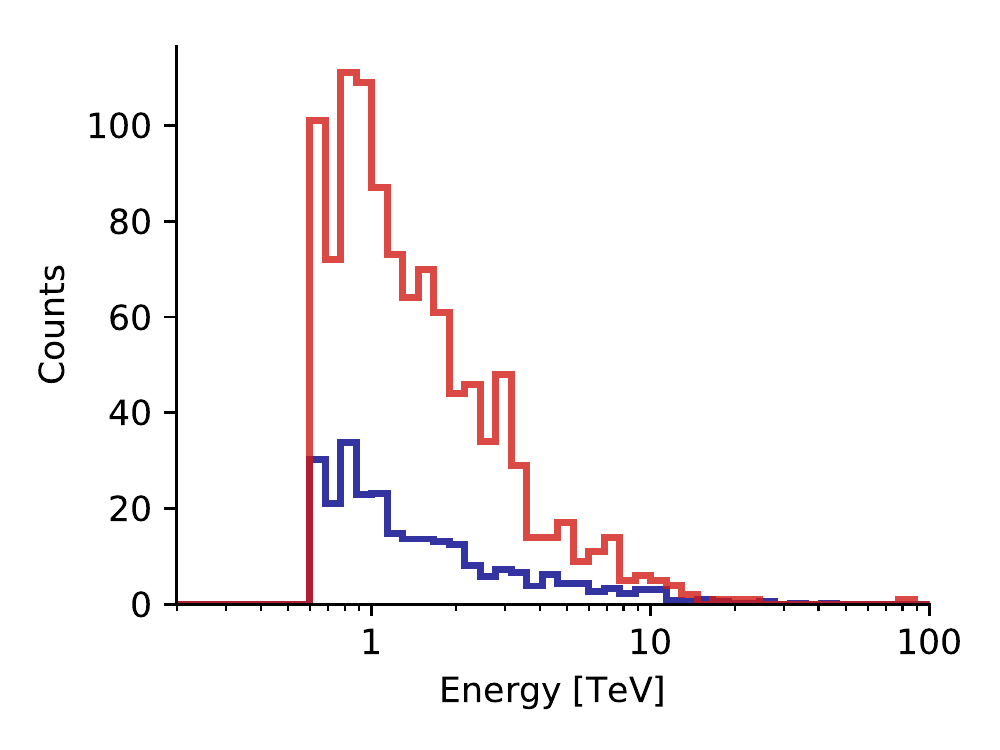}
\end{subfigure}
\caption{\crab \sourceImageDescr}
\label{fig:crab}
\end{figure}

This data release contains \dataTimeCrab~hours of observation (\dataRunsCrab~runs) of
the Crab nebula. The observations were taken in 2004 and are a very small subset
of the data used in the 2006 \hess\ paper on the Crab nebula
\cite{Aharonian:2006f}. Two of the runs are taken with the telescope pointing
with an offset of 0.5 deg from the source position, two runs with an offset of
1.5 deg. The data set is illustrated in Figure~\ref{fig:crab}. It contains $\sim
1000$ gamma rays, with a significant signal from energy threshold at $\sim
600$~GeV up to $\sim 10$~TeV. The energy threshold for this source is high
because the observations were taken at a high zenith angle (45-48~deg).

The Crab nebula was chosen for this data release because it is possibly the most
well-known and studied gamma-ray source. There is no variability in this data
set and the small size of the dataset and the low precision of the IRFs do not
allow for a precision measurement as recently done for the extension in
\cite{2017arXiv170704196H}.

\clearpage
\section{\pks}
\label{sec:ds:pks}

Two different sets of data are presented for the extra-galactic source \pks, an
active galactic nucleus (AGN) with bright and highly variable TeV emission. The
motivation to include these datasets in this data release was to have a variable
source. It is point-like with a known position, i.e. studies of this source will
focus on variability and spectrum.

\begin{figure}[h] \centering
\begin{subfigure}{.44\textwidth}
 \includegraphics[height=6cm]{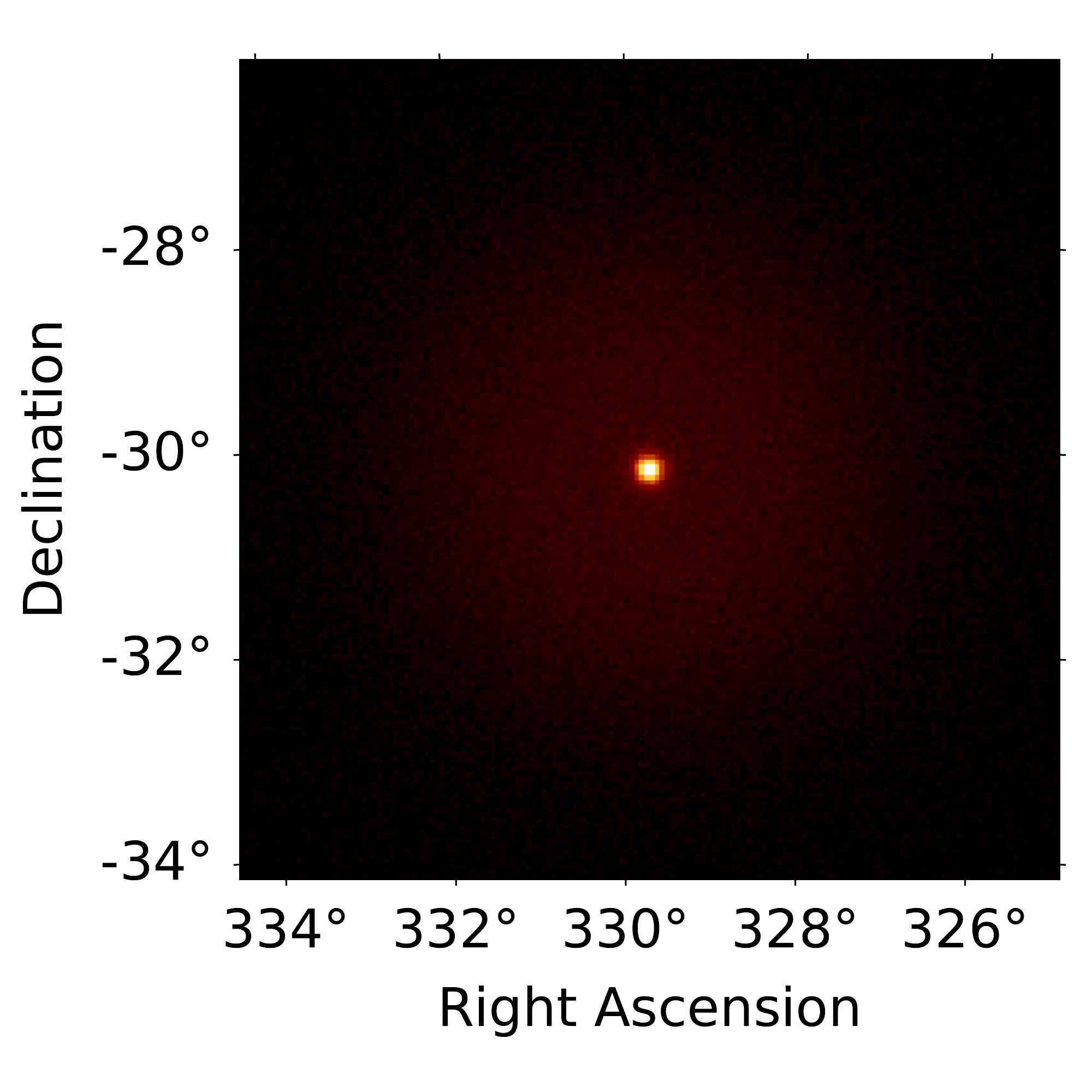}
\end{subfigure}
\begin{subfigure}{.55\textwidth}
  \includegraphics[height=6cm]{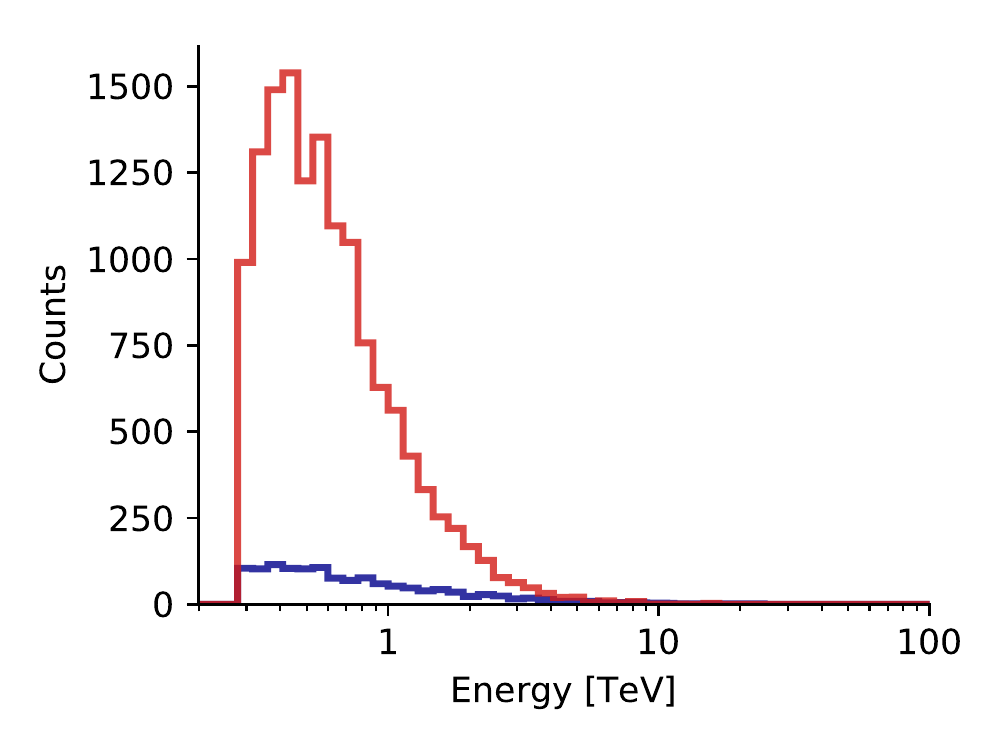}
\end{subfigure}
\caption{\pks (flare) \sourceImageDescr}
\label{fig:pks_flare}
\end{figure}

The \textbf{\pks (flare)} data set (see Figure~\ref{fig:pks_flare}) contains
\dataTimePksFlare~hours of observation (\dataRunsPksFlare~runs) from the nights
of July 29 and 30, 2006 (around MJD~53946), when the source underwent a major
gamma-ray outburst during its high-activity state of summer 2006. This \hess\
dataset as well as simultaneous observations with the Chandra satellite were
previously published in \cite{2009ApJ...696L.150A, 2009AaA...502..749A,
2012AaA...539A.149H}. All data were taken at an offset of 0.5~deg, spanning a
zenith angle range of 7-50~deg. The source was very bright and variable, the
total excess in this dataset is $\sim 21,000$ gamma rays.

\begin{figure}[h] \centering
\begin{subfigure}{.44\textwidth}
 \includegraphics[height=6cm]{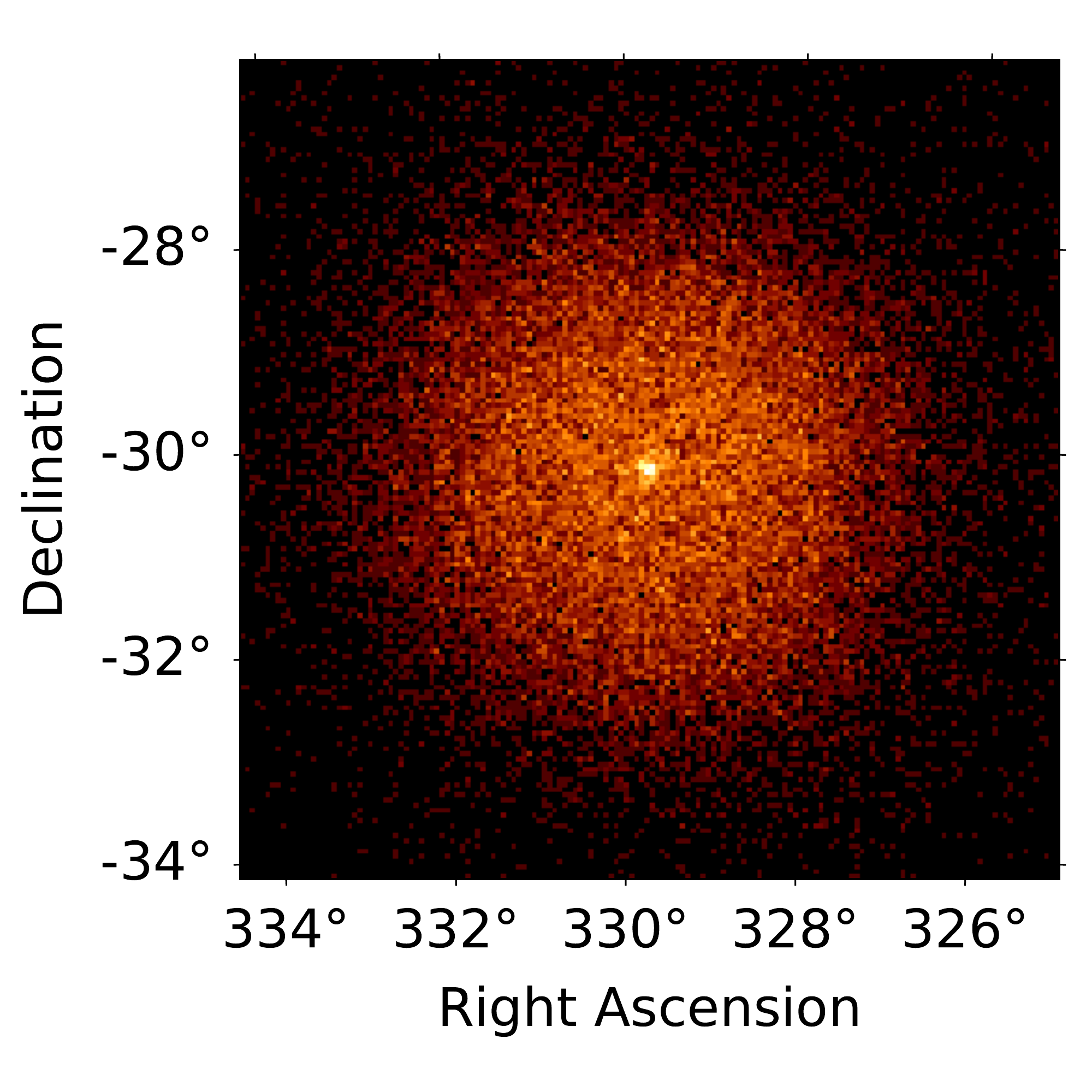}
\end{subfigure}
\begin{subfigure}{.55\textwidth}
  \includegraphics[height=6cm]{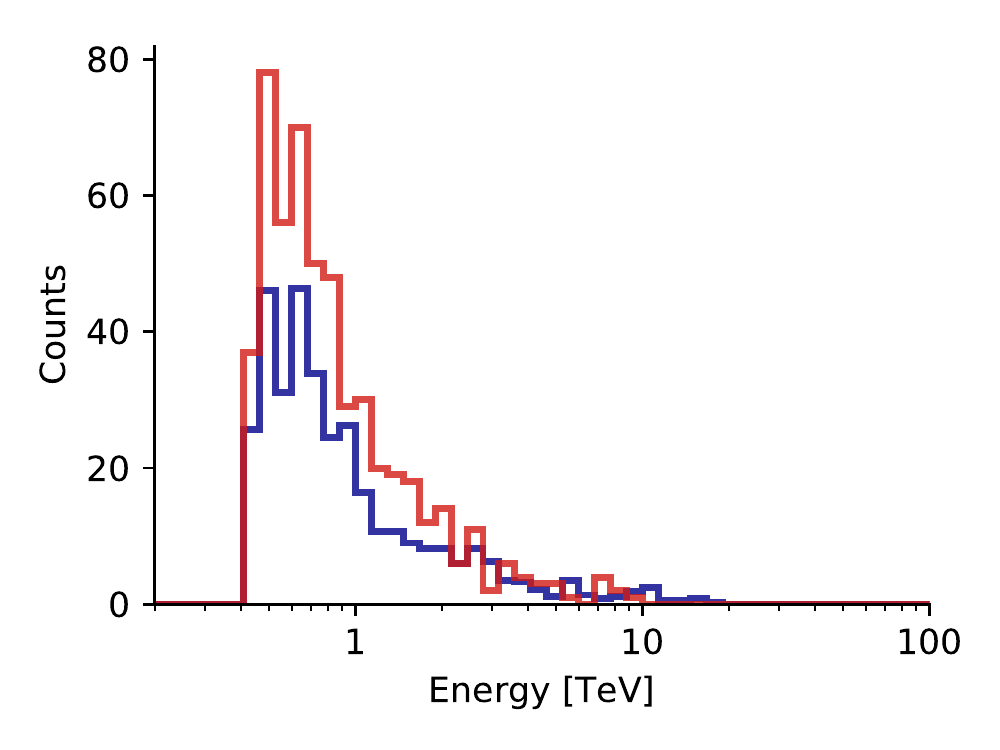}
\end{subfigure}
\caption{\pks (steady) \sourceImageDescr}
\label{fig:pks_steady}
\end{figure}

The \textbf{\pks (steady)} data set (see Figure~\ref{fig:pks_steady}) contains
\dataTimePksSteady~hours of observation (\dataRunsPksSteady~runs) from 2008,
taken at an offset of 0.5~deg and zenith angle of 23-37~deg, with an excess of
$\sim 400$ gamma rays.

\clearpage
\section{\msh}
\label{sec:ds:msh}

The supernova remnant \msh is a complex object with an unusual morphology. It
contains the pulsar PSR~B1509$-$58 and an extended, asymmetric pulsar wind nebula
that has been observed at X-ray energies by ROSAT, as well as more recently at
high angular resolution by Chandra. At TeV energies, \hess\ has also observed a
small, but clearly extended and elongated source \cite{Aharonian:2005c,
2017arXiv170901422T}. The TeV emission is thought to come from the pulsar wind
nebula, rather than from the pulsar or the supernova remnant.

\begin{figure}[h] \centering
\begin{subfigure}{.44\textwidth}
 \includegraphics[height=6cm]{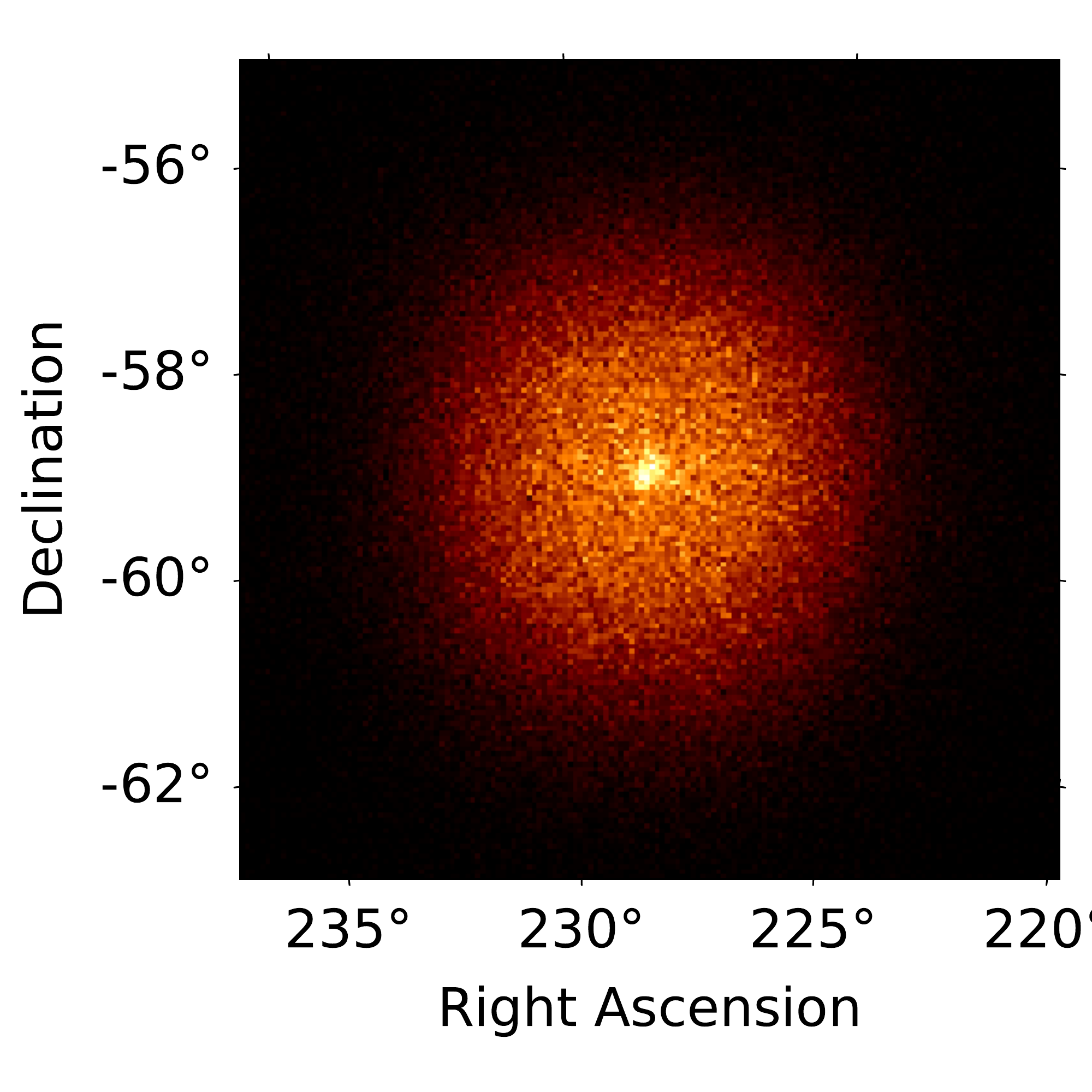}
\end{subfigure}
\begin{subfigure}{.55\textwidth}
  \includegraphics[height=6cm]{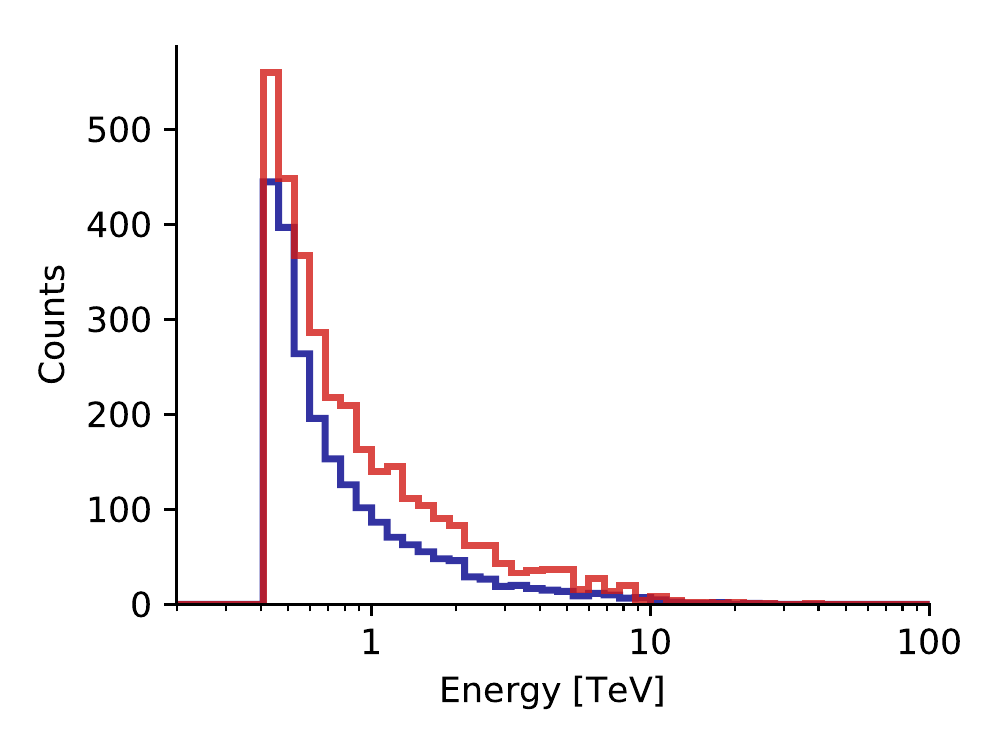}
\end{subfigure}
\caption{\msh \sourceImageDescr}
\label{fig:msh}
\end{figure}

This data release contains \dataTimeMsh~hours of observation (\dataRunsMsh~runs)
of \msh from 2004, a small subset of the data from the first \hess publication
on this source \cite{Aharonian:2005c}. All observations were taken at an offset
of 0.5~deg, at a zenith angle of 35-40~deg. The data set (see
Figure~\ref{fig:msh}) contains $\sim 1500$ gamma rays, with a significant signal
from energy threshold at $\sim 400$~GeV up to $\sim 10$~TeV.

The motivation to include this source in the data release was to have a small
extended source that allows morphology studies, i.e. measuring the source
position, extension and elongation. There are other TeV sources in the field of
view (HESS J1503-582, HESS J1457-593 and HESS J1458-608), but they are at an
offset of more than one degree and fainter than \msh, so obtaining good results
for \msh is possible without modeling those other sources.

\clearpage
\section{\rxj}
\label{sec:ds:rxj}

The supernova remnant \rxj is one of the largest ($\sim 1$~deg diameter) and
brightest TeV sources. It was selected for this data release as an example of a
very extended source with a complex morphology. As shown in previous \hess\
publications (\cite{Aharonian:2006e}, \cite{Aharonian:2007c},
\cite{2016arXiv160908671H}), gamma-ray emission is found all throughout the
shell-type supernova remnant, at varying levels of intensity.

\begin{figure}[h] \centering
\begin{subfigure}{.44\textwidth}
 \includegraphics[height=6cm]{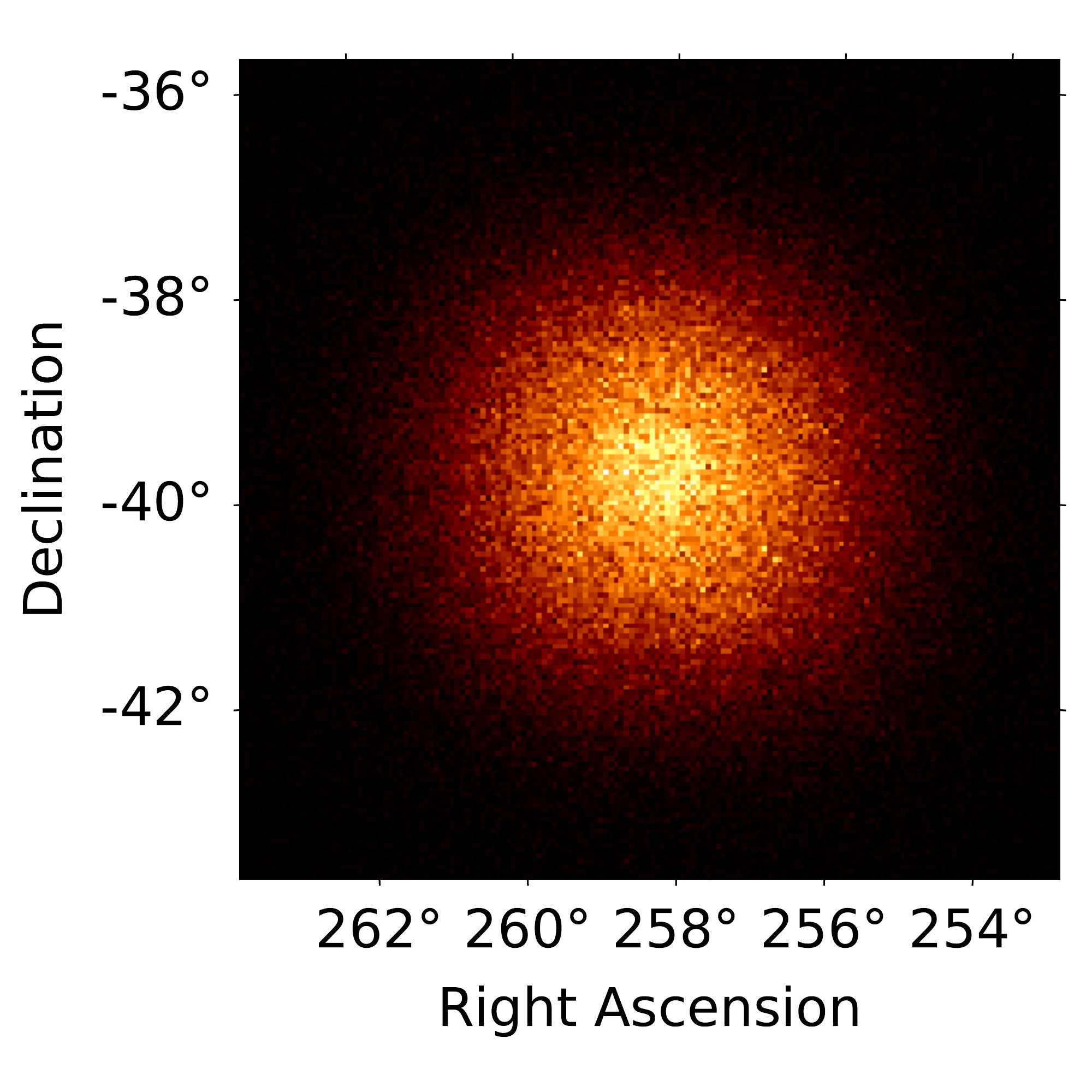}
\end{subfigure}
\begin{subfigure}{.55\textwidth}
  \includegraphics[height=6cm]{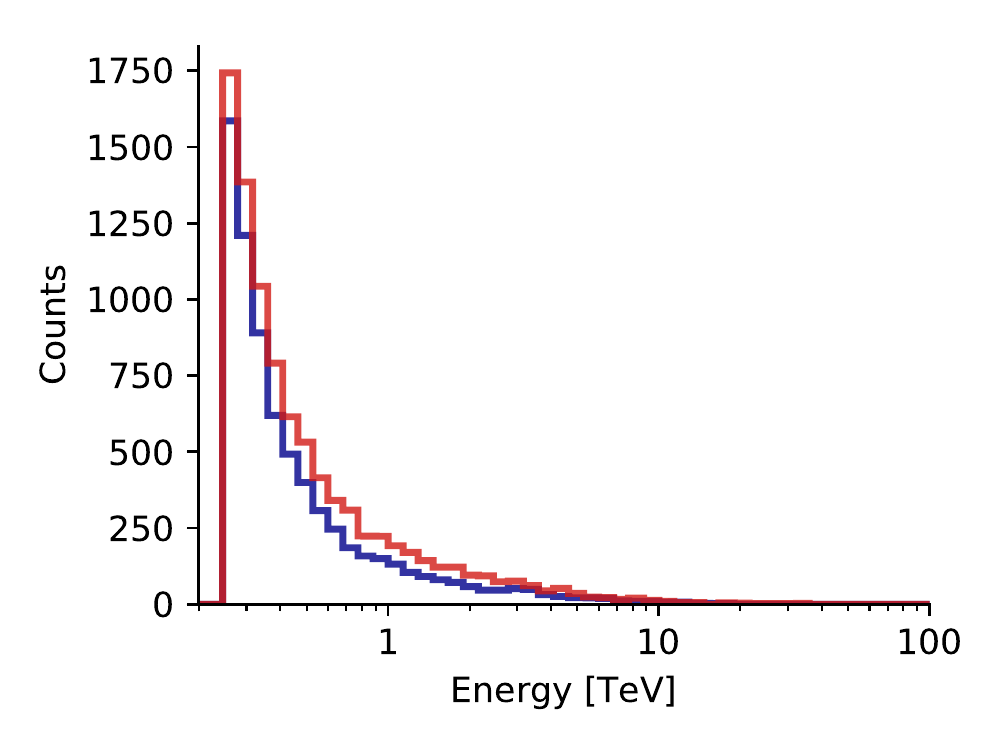}
\end{subfigure}
\caption{\rxj \sourceImageDescr}
\label{fig:rxj}
\end{figure}

This data release contains \dataTimeRxj~hours of observation (\dataRunsRxj~runs)
of \rxj from 2004, a subset of the data used in early \hess publications on this
source (\cite{Aharonian:2006e}, \cite{Aharonian:2007c}). Most observations were
taken at an offset of 0.7~deg from the center of the SNR (three observations
were pointing at the SNR center), at a zenith angle of 16-26~deg. The data set
(see Figure~\ref{fig:rxj}) contains $\sim 3600$ gamma rays, with a significant
signal from energy threshold at $\sim 250$~GeV up to $\sim 10$~TeV.

\clearpage
\section{Off runs}
\label{sec:ds:off}

Modeling the gamma-like hadronic background is perhaps the most difficult aspect
for many IACT data analyses. Background is usually estimated from real data (not
Monte-Carlo simulations), either from the mostly empty parts within the field of
view of a given run, or from other so called ``off'' observations of mostly
empty fields of view \cite{2007A&A...466.1219B, 2014A&A...568A.117F}. For the
\hess experiment there exists an ``off run list'' consisting mostly of
observations that did not result in a detection (and a small fraction of
dedicated off observations), that is used in \hess to construct background
models (usually by grouping in zenith angle bins).

Here we release a very small subset of \dataTimeOff~hours (\dataRunsOff~runs) of
the \hess-internal ``off run list'' that contains thousands of runs. It can be used as
a small example dataset to study \hess background, or to develop codes and
methods to create background models. However, we note that this is limited due
to the small size of this dataset, e.g. template background models that
represent the spatial shape and / or spectrum will be noisy due to the small
number of events available.

\chapterimage{layout/chapter_head9_small}
\chapter{Data files}
\label{sec:df}

\section{Overview}
\label{sec:df:overview}

The release notes document \path{hess_dl3_dr1.pdf} (the one you are reading at
the moment) is available as a separate file from the data release webpage.

The data from this release is contained in a gzipped tarball with filename
\path{hess_dl3_dr1.tar.gz}. To extract the content of the tarball, the following
commmand can be used:
\begin{verbatim}
  tar zxf hess_dl3_dr1.tar.gz
\end{verbatim}
This will result in a directory called \path{hess_dl3_dr1} with the
sub-directories and files shown in Figure~\vref{fig:dir}. The total size of the
files in this data release is \dataSizeFitsDataAll~MB.

The \path{README.txt} file contains a brief description of the data release, as
well as the terms of use.

The rest of the files are gzipped FITS files. FITS is a standard data exchange
and archival format in astronomy. It supports the storage of multiple header
data units (HDUs) in one FITS file. We store all data in binary table (BINTABLE)
HDUs.

The observation index table in \path{obs-index.fits.gz} and HDU index table in
\path{hdu-index.fits.gz} can be used to select and load data, they are
described in Section~\ref{sec:df:index}.

The data for each observation run is contained in a single FITS with name
\path{hess_dl3_dr1_obs_id_NNNNNN.fits.gz}, where \verb=NNNNNN= is the
\verb=OBS_ID= number of the run. For each run there are five HDUs (names:
\verb=EVENTS=, \verb=GTI=, \verb=AEFF=, \verb=EDISP= and \verb=PSF=) that are
summarised in Table~\ref{tab:hdus} and described in the remaining sections of
this chapter.

A detailed data format specification is available separately in the ``Data
formats for gamma-ray astronomy'' document version \gadfVersion, which is
available on Github\footnote{\urlGadfGithub}, Readthedocs\footnote{\urlGadfDocs}
and archived on Zenodo~\cite{deil_christoph_2018_1409831}. This chapter only
describes additional information that is specific to this \hess data release, such as
e.g.\ which IRF formats we use and what axis binnings. Further information on
how this DL3 FITS data was produced, as well as notes and caveats concerning the
IRFs, is given in Section~\ref{sec:notes:prod}.


\begin{figure}[t]
\centering

\begin{forest}
  for tree={
    font=\ttfamily,
    grow'=0,
    child anchor=west,
    parent anchor=south,
    anchor=west,
    calign=first,
    edge path={
      \noexpand\path [draw, \forestoption{edge}]
      (!u.south west) +(7.5pt,0) |- node[fill,inner sep=1.25pt] {} (.child anchor)\forestoption{edge label};
    },
    before typesetting nodes={
      if n=1
        {insert before={[,phantom]}}
        {}
    },
    fit=band,
    before computing xy={l=15pt},
  }
[
  [hess\_dl3\_dr1.pdf]
  [hess\_dl3\_dr1.tar.gz]
  [hess\_dl3\_dr1
    [README.txt]
    [hdu-index.fits.gz]
    [obs-index.fits.gz]
    [data
      [hess\_dl3\_dr1\_obs\_id\_NNNNNN.fits.gz]
      [hess\_dl3\_dr1\_obs\_id\_NNNNNN.fits.gz]
      [...]
    ]
  ]
]
\end{forest}

\caption[Directory structure and files]{
Directory structure and files in the release tarball. The
\texttt{hess\_dl3\_dr1.pdf} and \texttt{hess\_dl3\_dr1.tar.gz} files are
part of the data release. After downloading and extracting
\texttt{hess\_dl3\_dr1.tar.gz} via \texttt{tar xvf
hess\_dl3\_dr1.tar.gz} you will find the \texttt{hess\_dl3\_dr1} folder
and files as shown.
}
\label{fig:dir}
\end{figure}

\begin{table}[t]
\centering
\begin{tabular}{llllrrr}
HDU    & Description           & HDUCLAS4   &  Rows & Cols & Size (kB) \\
\midrule
EVENTS & Event parameters      &            &  \dataEventsCountMean & 5    & \dataSizeEvents \\
GTI    & Good time intervals   &            &  2    & 2    & \dataSizeGti \\
AEFF   & Effective area        & AEFF\_2D   &  1    & 5    & \dataSizeAeff \\
EDISP  & Energy dispersion     & EDISP\_2D  &  1    & 7    & \dataSizeEdisp \\
PSF    & Point spread function & PSF\_TABLE &  1    & 7    & \dataSizePsf \\
\end{tabular}
\caption[FITS HDU overview]{
FITS HDU overview. All HDUs are \texttt{BINTABLE} HDUs, all IRFs are
full-enclosure IRFs. Mean HDU size is given in kilo-bytes (kB). Number of
rows for EVENTS is the average, for the other tables it is always the same.
}
\label{tab:hdus}
\end{table}

\clearpage
\section{Index files}
\label{sec:df:index}

The observation index table in \path{obs-index.fits.gz} and HDU index table in
\path{hdu-index.fits.gz} can be used to select and load data. Their format is
described in the open specifications. We note that using these index files is
optional, with a small effort to select runs and declare the input data
correctly, it is possible to access and load the data with Gammapy or ctools
without using these index files.

The observation index table in \path{obs-index.fits.gz} has \dataRunsTotal~rows,
one row per observation run. The \dataObsIndexCols~columns list the
parameters that can be useful for run selection. The
\verb=OBS_ID= identifies the observation, and e.g.\ the \verb=RA_PNT= and
\verb=DEC_PNT= columns give the run pointing position. All of the information
shown in the Tables~\ref{tab:sources}, \ref{tab:obs_summary},
\ref{sec:appendix:obs_list} and in Figure~\ref{fig:ds:obs} is contained in the
observation index table. Most of the parameters for a given observation are also
contained in the \verb=EVENTS= FITS header under the same key name. A few more
columns have been added for convenience, e.g.\ the \verb=SAFE_ENERGY_LO= in
the observation table is taken from the \verb=LO_THRES= key in the \verb=AEFF=
FITS header. The \verb=TARGET_NAME= key was added that gives the observation
subset name shown in Table~\ref{tab:obs_summary}; this is the easiest way to
select e.g.\ all ``PKS 2155-304 (flare)'' or all ``Off data'' runs in this data
release.

The HDU index table in \path{hdu-index.fits.gz} can be used to locate and load
any FITS HDU. It has $5 \times \dataRunsTotal$ rows (one for each of the 5
different HDUs and each observation run) and \dataHduIndexCols~columns that give
the type, size and location of each HDU, i.e. the folder relative to the index
file, the filename and the HDU name.

\clearpage
\section{Events}
\label{sec:df:events}

The \hess data is given as an event list in a FITS HDU called \texttt{EVENTS}
for each observation, with columns \verb=EVENT_ID=, \verb=TIME=, \verb=RA=,
\verb=DEC= and \verb=ENERGY=. 

The \hess observatory location (usually not needed for science analysis) is:

\begin{verbatim}
    GEOLAT  =    -23.2717777777778 / latitude of observatory (deg)
    GEOLON  =     16.5002222222222 / longitude of observatory (deg)
    ALTITUDE=                1835. / altitude of observatory (m)
\end{verbatim}

The \hess reference time is defined following the FITS time standard, with the
following values:

\begin{verbatim}
    MJDREFI =                51910 / int part of reference MJD for times
    MJDREFF = 0.000742870370370241 / fractional part of MJDREF
    TIMEUNIT= 's       '           / time unit is seconds since MET start
    TIMESYS = 'TT      '           / Time system (TT=terrestrial time)
    TIMEREF = 'local   '           / local time reference
\end{verbatim}

In \hess, there is no unique EVENT ID. Instead, there are two numbers that
together uniquely identify an event within a given \verb=OBS_ID=, the so-called
bunch number (the result of how data acquisition works in \hess) and event
number within a bunch. To comply with the DL3 spec, which requires a unique
\verb=EVENT_ID= within a given \verb=OBS_ID=, we have decided to fill
\verb=EVENT_ID= as follows in the \hess FITS exporters:\\

\begin{verbatim}
    EVENT_ID = (BUNCH_ID_HESS << 32) | (EVENT_ID_HESS)
\end{verbatim}

\section{Good time intervals}
\label{sec:df:gti}

The good time interval table \texttt{GTI} is something that is commonly used in
high-energy missions since decades to declare the observation times
corresponding to the given events. For this \hess\ data we give it as well, even
though the GTI tables always consist of a single row giving the start and stop
time for each observation. The same information is already present in the header
of the \texttt{EVENTS} extension under the \texttt{TSTART} and \texttt{TSTOP}
keys, and in addition in a timestamp string format via the \texttt{DATE-OBS} and
\texttt{TIME-OBS} (start) and \texttt{DATE-END} and \texttt{TIME-END} (stop)
keys. To compute exposures and fluxes from the data released here, the
\verb=LIVETIME= header key in the \verb=EVENTS= HDU can be accessed, or
equivalently, the \verb=ONTIME= and dead time correction factor \verb=DEADC= from there
could be used, since\\

\begin{verbatim}
    ONTIME = TSTOP - TSTART
    LIVETIME = DEADC x ONTIME
\end{verbatim}

\clearpage
\section{Instrument responses}
\label{sec:df:irf}

This section contains general comments on the instrument response function (IRF)
information that is described in the following sections. For every observation
we assume that the instrument response function (IRF) is stable. This is an
approximation, in reality there is a small variation in response, mainly because
of the zenith angle variation during the run.

\vspace{.2cm}\noindent The response is stored in FITS HDUs, for the following quantities:

\begin{itemize}
\item \textbf{aeff}: Effective area in \verb=aeff_2d= format. See Section~\ref{sec:df:aeff}.
\item \textbf{edisp}: Energy dispersion in \verb=edisp_2d= format. See Section~\ref{sec:df:edisp}.
\item \textbf{psf}: Point spread function in \verb=psf_table= format. See Section~\ref{sec:df:psf}.
\end{itemize}

We note that the MC statistics for the IRFs used here is high, even at high
energies. The Poisson noise and re-sampling artifacts are relatively low, and
the dependence of IRFs on parameters like energy or offset is usually smooth. In
practice this means that science tools can directly use the IRFs using linear
interpolation or even nearest-bin queries and obtain good results.

\vspace{.2cm}\noindent That said, we note that no quantitative IRF error is
given. When analyzing this data, please note the following caveats:

\begin{itemize}

\item Some instrumental effects (e.g. broken pixels in the camera images) are
known to broaden the gamma-ray PSF, yet are not taken into account in the MC
point-source simulations here. No evaluation of the PSF systematics and
precision is given for this PSF.

\item The assumed IRFs in the data release are computed for the mean zenith angle 
during the run, where in reality the zenith angle varies somewhat during the 
run and across the field of view. 

\item Similarly, IRFs are computed from point-source simulations at fixed zenith,
azimuth and field-of-view offset angles, so for any given source position some
interpolation error results. This can be a problem in particular close to the
energy threshold, which is a function of, e.g., the zenith angle.

\item The offset binning has been chosen to reflect the \hess simulation of
IRFs, which is carried out at six different offset angles, namely 0.0, 0.5, 1.0,
1.5, 2.0, and 2.5~deg. Since the IRFs are computed using point-source
simulations at fixed offsets, the bins are defined such that their low and high
edges are identical, and equal to the offset angle that was simulated (e.g. the
first bin has edges (0~deg, 0~deg)).

\end{itemize}

The IRF uncertainties translate into systematic errors on high-level analysis
results. For previous \hess\ publications, our knowledge of the whole
instrumental chain, the uncertainties of the Monte Carlo simulations and of the
analysis chain (calibration, reconstruction and discrimination) lead to an
estimation of the systematic errors of 20\% on the flux and 0.1 on the spectral
index for a bright isolated point source \cite{Aharonian:2006f}, and up to 30\%
and 0.2 respectively or more for extended sources in the Galactic plane
\cite{hgps}. The systematic error of the reconstructed source location is less
than 20~arcsec in RA and Dec \cite{Braun:2007}. For this FITS dataset, no
evaluation of systematic errors has been performed yet.

\clearpage
\section{Effective area}
\label{sec:df:aeff}

\begin{figure}[h]
\centering
\includegraphics[width=\linewidth]{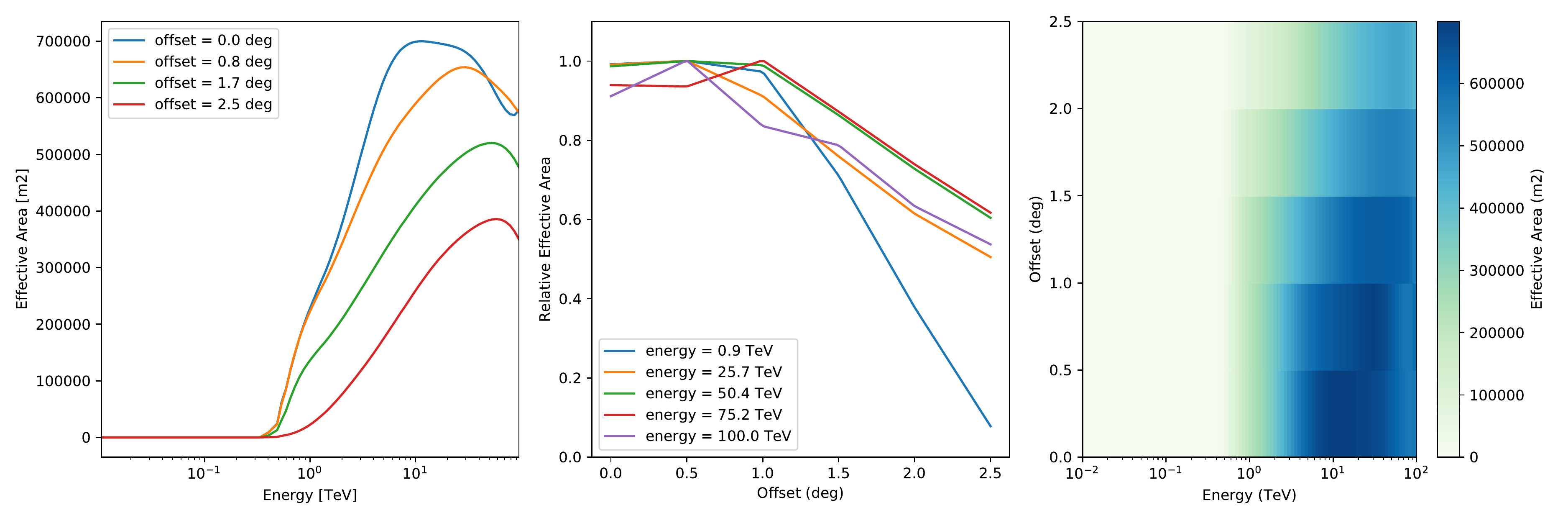}
\caption[Effective area]{
Effective area for observation \texttt{OBS\_ID=23523} at the Crab nebula
position (\crabIrfDescr).
}
\label{fig:aeff}
\end{figure}

Effective areas are stored in the \verb=aeff_2d= format and are
illustrated in Figure~\vref{fig:aeff}.

\vspace{.2cm}\noindent The effective area IRF has two axes:

\begin{itemize}

\item The field of view offset axis has bins located at 0, 0.5, 1.0, 1.5, 2.0
and 2.5 deg. This is identical for all IRFs and explained in
Section~\ref{sec:notes:prod:irfs}.

\item The energy binning is equally spaced in the logarithm of the energy, with
96 bins from 0.01~TeV to 100~TeV (24 bins per decade in energy). This energy
binning is fine enough so that there should be no effects due to interpolation
performed by science tools when evaluating the data.

\end{itemize}

\vspace{.2cm}\noindent For technical reasons, the curves have been smoothed
by fitting a high-degree polynomial function to the simulated histogram. This
fit can sometimes diverge in the last few bins at the highest energies (due
to missing simulation data at even higher energies). However, it has been
verified that the fit is still in good agreement with the underlying histogram
within its statistical uncertainties in all cases.

\clearpage
\section{Energy dispersion}
\label{sec:df:edisp}

\begin{figure}[h]
\centering
\includegraphics[width=\linewidth]{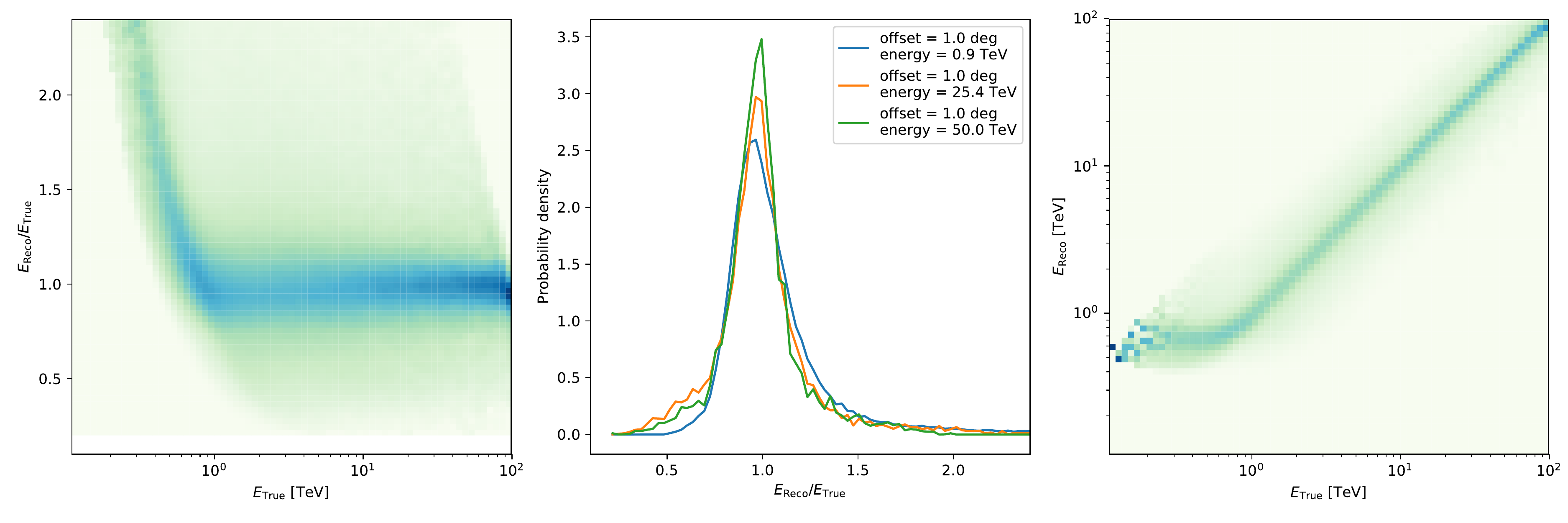}
\caption[Energy dispersion]{
Energy dispersion for observation \texttt{OBS\_ID=23523} at the Crab nebula
position (\crabIrfDescr).
}
\label{fig:edisp}
\end{figure}

Energy dispersion is stored in the \verb=edisp_2d= format and is
illustrated in Figure~\vref{fig:edisp}.

\vspace{.2cm}\noindent The energy dispersion IRF has three axes:

\begin{itemize}

\item The field of view offset axis has bins located at 0, 0.5, 1.0, 1.5, 2.0
and 2.5 deg. This is identical for all IRFs and explained in
Section~\ref{sec:notes:prod:irfs}.

\item The energy axis has 96~bins, logarithmically spaced between 0.01~TeV and
100~TeV.

\item The \verb=MIGRA= axis (defined as reconstructed over true energy ratio)
has a linear binning with a bin width of 0.03, ranging from 0.2 and 5.0 (160
bins). This bin width of 3\% energy resolution is good enough to capture the
shape of the \hess energy dispersion, which has a width of roughly 15\%.

\end{itemize}

No smoothing was applied.

\clearpage
\section{Point spread function}
\label{sec:df:psf}

\begin{figure}[h]
\centering
\includegraphics[width=\linewidth]{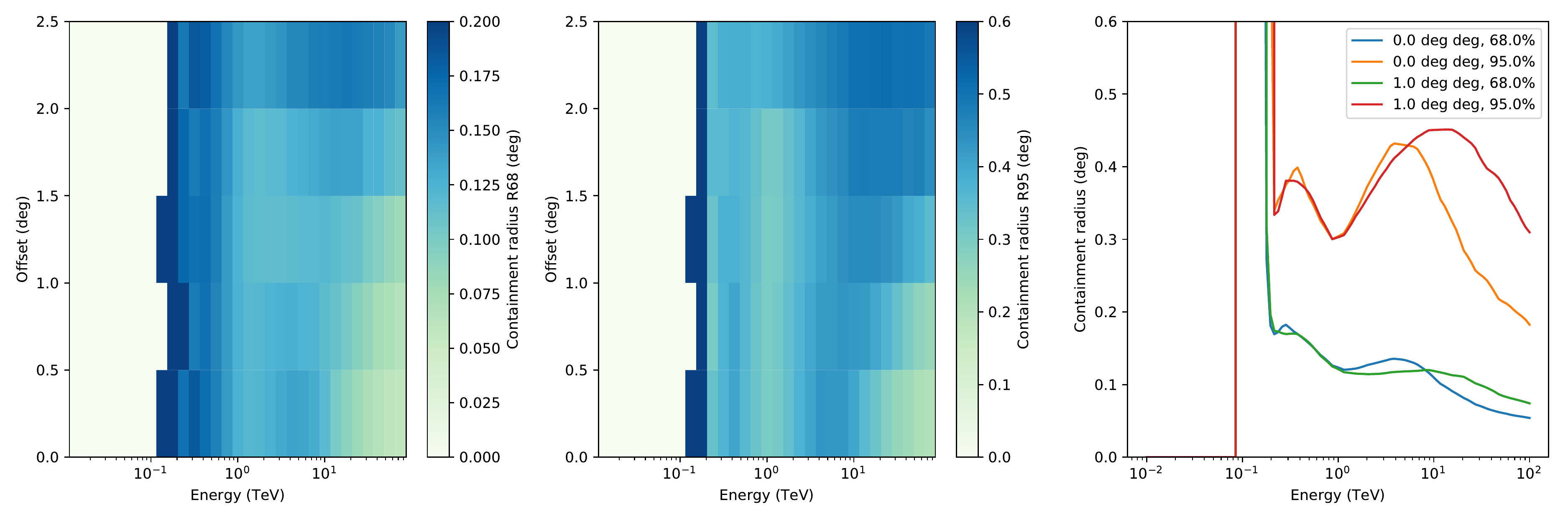}
\caption[Point spread function]{
Point spread function for observation \texttt{OBS\_ID=23523} at the Crab nebula
position (\crabIrfDescr).
}
\label{fig:psf}
\end{figure}

The point spread function is stored in the \verb=psf_table= format and is
illustrated in Figure~\vref{fig:psf}. It is assumed to be radially symmetric.

\vspace{.2cm}\noindent The point spread function IRF has three axes:

\begin{itemize}

\item The field of view offset axis has bins located at 0, 0.5, 1.0, 1.5, 2.0
and 2.5 deg. This is identical for all IRFs and explained in
Section~\ref{sec:notes:prod:irfs}.

\item The energy binning is equally spaced in the logarithm of the energy, with
32 bins from 0.01~TeV to 100~TeV.

\item The binning of the radial parameter \verb=RAD= of the point spread
function is equally spaced in angle-squared up to 0.1~deg. Further bins are
equally spaced in square-root of the angle. The PSF is stored for
radial offsets from 0~deg to 0.67~deg in the file (144 bins in total).

\end{itemize}

This binning in \verb=RAD= has been chosen in order to preserve all information
about the shape of the point spread function in its core region that is
available from the simulated \hess IRF. At positions further away from the
core the binning can be more coarse because the point spread function changes
very slowly with increasing distance to the core.

No smoothing was applied.

\chapterimage{layout/chapter_head4_small}
\chapter{Notes}
\label{sec:notes}

This section contains additional notes: Section~\ref{sec:notes:prod} describes
in detail how these FITS data were produced, mostly focused on information and
caveats for the instrument response functions. Section~\ref{sec:notes:ana}
gives information how to analyse this data.

\section{\hess DL3 FITS production}
\label{sec:notes:prod}

This section contains detailed information about how the DL3 data were produced.

It comprises software and version numbers, but also general information on data
calibration, event reconstruction, gamma-hadron separation and IRF production
that are important to explain the features visible in the event parameter
distributions or IRFs.

\subsection{Overview}
\label{sec:notes:prod:overview}

To produce DL3 data, the following main steps are performed: calibration, event
reconstruction, gamma-hadron separation, data quality selection. For this data
release we decided to use what is known in \hess as ``Heidelberg calibration'',
event stereo ``Hillas reconstruction'', ``standard cuts'' with full enclosure
effective areas and ``spectral data quality selection''. This is described in
detail in the ``Observations of the Crab nebula with \hess'' paper from 2006
\cite{Aharonian:2006f}. All data was taken with the \hessOne array, with all 4
telescopes participating in each observation, and at least two telescopes
triggering for each event.

We would like to note that Hillas reconstruction and standard cuts do not
represent the state of the art for \hess analysis. All recent and ongoing \hess
publications use better, more sensitive methods, e.g. multivariate gamma-hadron
separation such as \cite{Ohm:2009} or \cite{Becherini:2011} or air shower
template-based event reconstruction such as \cite{2009APh....32..231D} or
\cite{2014APh....56...26P}. As stated in Section~\ref{sec:context}, the goal of
this data release is to test data models, formats and science tools, not to
offer the most sensitive \hess analysis. The main reason we chose Hillas
standard cuts is that the data and IRFs are stable (updated last in 2011),
well-tested and robust. The more modern analyses with better background
rejection and sensitivity typically are less stable (e.g. with respect to small
calibration issues), and exhibit features that complicate the analysis, such as
steps in energy for effective area and event distributions due to training of
gamma-hadron separation in energy bands.

This FITS dataset was produced using DST version \dstVersion, and the HAP
analysis configuration \verb=std_fullEnclosure_fits_release=. The software used
was the \verb=hess-data-fits-export= batch script, which calls \verb=hap= to
export the \verb=EVENT= data and \verb=hap-to-irf= to export the IRFs, using the
software version as of \hapVersionDate. The script used Astropy~\astropyVersion
and Gammapy~\gammapyVersion to process and check the data.

\subsection{IRFs}
\label{sec:notes:prod:irfs}

The effective area, energy dispersion and PSF IRFs were computed from point-source
Monte Carlo (MC) simulations with a power-law energy spectrum (spectral index~$-2$)
and a grid of values in the following parameters:

\begin{itemize}

\item Zenith angles: 0, 10, 20, 30, 40, 45, 50, 55, 60, 63, 65 deg

\item Offset angles: 0.0, 0.5, 1.0, 1.5, 2.0, 2.5 deg

\item Azimuth angles: 0, 180 deg (north and south)

\item Muon efficiency configs: 100, 101, 102, 103, 104, 105

\end{itemize}
The muon efficiency configurations correspond to different intervals in time with
varying optical efficiency of the telescopes. Their labels are arbitrary and carry
no physical information.

The IRFs for a given run were computed by interpolating between these parameters
/ configurations. For the pointing position, the average zenith and azimuth angles 
of the events recorded during the observation were used. The \hess lookup
production, and the computation of ``effective'' IRFs for a given run as
performed here, are complicated in detail, and described in the \hess-internal
note \cite{lookups_internal_note}.

\section{Analysis recommendations}
\label{sec:notes:ana}

This section contains some information how to analyse the data presented here.

\subsection{Tools}
\label{sec:notes:ana:tools}

The data in this release are in a standard, well-documented format (for its
definition, see \urlGadfDocs and \cite{2016arXiv161001884D}). At the time of this
data release, this format is supported by two open-source science tool packages:
Gammapy (see \urlGammapy and \cite{2015arXiv150907408D, 2017arXiv170901751D}),
currently at version~\gammapyVersion and ctools (see \urlCtools and \cite{2016AnA...593A...1K}),
currently at version 1.5.

We do not give any tutorial or present any results here for the analysis of this data.
A future publication with a detailed analysis and comparison of \hess data with
the internal tools, Gammapy and ctools is in preparation. Tutorials for Gammapy
or ctools or other science tool packages analysing this data are very welcome,
but will be left to the science tool teams to create and maintain.

\subsection{Safe cuts}
\label{sec:notes:ana:cuts}

When analysing these data, it is recommended to apply a safe energy
and maximum field of view (FOV) cut.

The \verb=AEFF_2D= table includes a keyword \verb=LO_THRESH=, which denotes the
lower ``safe'' energy threshold for the observation. The threshold was computed
as the energy at which the energy bias equals 10\%, at an offset of 1~degree in
the field of view. Figure~\ref{fig:ds:obs} shows the distribution of ``safe''
energy threshold values for the observation runs in this data release. It varies
between 200~GeV and 1~TeV. Higher zenith-angle runs have a larger energy
threshold, and to a lesser degree also later runs have a higher energy
threshold, because the optical efficiency of the telescopes decreases over time.

For analyses based on these \hess data, we recommended to apply a minimum energy cut
matching this safe energy threshold when performing an analysis,
because the responses at lower energies are unreliable
(all of them: effective area, energy dispersion and point spread function).

It is also recommended to apply a maximum field of view offset cut of 2~degrees,
because at larger offsets the responses are unreliable.

\subsection{Background modeling}
\label{sec:notes:ana:bg}

This data release does not contain background models. To analyse it, we recommend
you use background modeling techniques that estimate the background from the
event data such as the reflected background method to derive spectra or the ring
background method without an offset acceptance correction, or use the off run
data given to estimate the background via the on-off method, or by generating a
background template model from the off runs. A good reference for background
estimation methods in VHE gamma-ray astronomy is \cite{2007A&A...466.1219B}.

\chapterimage{layout/chapter_head10_small}
\chapter{Acknowledgements}
\label{sec:ack}

We would like to thank everyone that has contributed to the open data for
gamma-ray astronomy effort at \urlGadfDocs.

In the preparation of this data release, we have made use of the SIMBAD
database, operated at CDS, Strasbourg, France \cite{2000AnAS..143....9W} as well
as NASA's Astrophysics Data System. The data processing was done with the \hess
software and Python scripts using Astropy, a community-developed core Python
package for Astronomy \cite{2013AnA...558A..33A, 2018arXiv180102634P}, and
Gammapy \cite{2015arXiv150907408D, 2017arXiv170901751D}.

The pictures used throughout this document were taken by Dalibor Nedbal,
Clementina Medina and Christian F\"ohr.

The support of the Namibian authorities and of the University of Namibia in
facilitating the construction and operation of H.E.S.S. is gratefully
acknowledged, as is the support by the German Ministry for Education and
Research (BMBF), the Max Planck Society, the German Research Foundation (DFG),
the Alexander von Humboldt Foundation, the Deutsche Forschungsgemeinschaft, the
French Ministry for Research, the CNRS-IN2P3 and the Astroparticle
Interdisciplinary Programme of the CNRS, the U.K. Science and Technology
Facilities Council (STFC), the IPNP of the Charles University, the Czech Science
Foundation, the Polish National Science Centre, the South African Department of
Science and Technology and National Research Foundation, the University of
Namibia, the National Commission on Research, Science \& Technology of Namibia
(NCRST), the Innsbruck University, the Austrian Science Fund (FWF), and the
Austrian Federal Ministry for Science, Research and Economy, the University of
Adelaide and the Australian Research Council, the Japan Society for the
Promotion of Science and by the University of Amsterdam. We appreciate the
excellent work of the technical support staff in Berlin, Durham, Hamburg,
Heidelberg, Palaiseau, Paris, Saclay, and in Namibia in the construction and
operation of the equipment. This work benefited from services provided by the
H.E.S.S. Virtual Organisation, supported by the national resource providers of
the EGI Federation.

\chapterimage{layout/chapter_head6_small}
\chapter*{References}

\addcontentsline{toc}{chapter}{\textcolor{hesscolor}{References}}
\printbibliography[heading=bibempty]

\chapterimage{layout/chapter_head11_small}
\begin{appendix}
\chapter{Appendix}
\label{sec:appendix}

\section{Acronyms}
\label{sec:appendix:acronyms}

The following abbreviations are used throughout this document:

\begin{acronym}[AAAAAA]
  \acro{VHE}{Very high energy (above $\sim$100~GeV)}
  \acro{DL3}{Data level 3}
  \acro{GTI}{Good time interval}
  \acro{FITS}{Flexible Image Transport System}
  \acro{HDU}{Header data unit of a FITS file}
  \acro{Run}{Alternative term for ``observation'', commonly used for IACT data}
  \acro{IACT}{Imaging atmospheric Cherenkov telescope}
  \acro{hess}[\hess]{High energy stereoscopic system}
  \acro{CTA}{Cherenkov telescope array}
  \acro{PWN}{Pulsar wind nebula}
  \acro{SNR}{Supernova remnant}
  \acro{AGN}{Active Galactic nucleus}
  \acro{MC}{Monte Carlo}
  \acro{HAP}{\hess analysis program; one of the \hess-internal analysis codes}
  \acro{FOV}{Field of view}
  \acro{offset}{Usually refers to offset in the field of view, i.e. with respect to the pointing position}
  \acro{IRF}{Instrument response function}
  \acro{AEFF}{Effective area}
  \acro{EDISP}{Energy dispersion}
  \acro{PSF}{Point spread function}
  \acro{obsid}[OBS\_ID]{Observation identifier, a.k.a. run number}
\end{acronym}

\newpage

\section{Full table of observations}
\label{sec:appendix:obs_list}

\begin{center}
\begin{longtable}{llrrrrrrl}
\caption[Full table of observations]{
Full table of observations. \texttt{OBS\_ID} is the observation identifier
(a.k.a. ``run number''). \texttt{Duration} is the observation time (not deadtime
corrected). The run pointing position is given in equatorial coordinates
(\texttt{RA} and \texttt{DEC}) as well as Galactic coordinates (\texttt{GLON}
and \texttt{GLAT}). \texttt{Offset} is the observation target object offset in
the field of view, taken as \texttt{RA\_OBJ} and \texttt{DEC\_OBJ} from the FITS
event list header, which is filled with the proposed target of observation from
the \hess database on FITS export. \texttt{Zenith} is the zenith angle of the
observation at mid run time. A summary of the available observations grouped by
sources is given in Table~\vref{tab:obs_summary}.
}\\

\hline
OBS\_ID & Date & Duration     & RA           & DEC          & GLON          & GLAT          & Offset       & Zenith       \\
        &      & \textit{min} & \textit{deg} & \textit{deg} & \textit{deg}  & \textit{deg}  & \textit{deg} & \textit{deg} \\
\hline
\endfirsthead
\multicolumn{7}{l}{\tablename\ \thetable\ -- \textit{Continued from previous page}} \\
\hline
OBS\_ID & Date & Duration     & RA           & DEC          & GLON          & GLAT          & Offset       & Zenith       \\
        &      & \textit{min} & \textit{deg} & \textit{deg} & \textit{deg}  & \textit{deg}  & \textit{deg} & \textit{deg} \\
\hline
\endhead
\hline
\multicolumn{7}{l}{\tablename\ \thetable\ -- \textit{Continued on next page}} \\
\endfoot
\endlastfoot

\\ \multicolumn{7}{l}{\textit{\underline{\crab observations (\dataRunsCrab~runs)}}}\\[3pt]
 23523 & 2004-12-04   &  28.1 &    83.6 &    21.5 &   185.0 &    $-$6.1 & 0.5 & 48 \\
 23526 & 2004-12-04   &  28.1 &    83.6 &    22.5 &   184.1 &    $-$5.5 & 0.5 & 45 \\
 23559 & 2004-12-06   &  28.1 &    85.3 &    22.0 &   185.4 &    $-$4.5 & 1.5 & 45 \\
 23592 & 2004-12-08   &  28.1 &    82.0 &    22.0 &   183.7 &    $-$7.0 & 1.5 & 48 \\

\\ \multicolumn{7}{l}{\textit{\underline{\pks (flare) observations (\dataRunsPksFlare~runs)}}} \\[3pt]
 33787 & 2006-07-29   &  28.1 &   329.7 &   $-$29.7 &    18.5 &   $-$52.2 & 0.5 & 50 \\
 33788 & 2006-07-29   &  28.1 &   329.1 &   $-$30.2 &    17.6 &   $-$51.7 & 0.5 & 43 \\
 33789 & 2006-07-29   &  28.1 &   330.3 &   $-$30.2 &    17.8 &   $-$52.7 & 0.5 & 37 \\
 33790 & 2006-07-29   &  28.2 &   329.7 &   $-$30.7 &    16.9 &   $-$52.3 & 0.5 & 30 \\
 33791 & 2006-07-29   &  28.1 &   329.7 &   $-$29.7 &    18.5 &   $-$52.2 & 0.5 & 24 \\
 33792 & 2006-07-29   &  28.1 &   329.1 &   $-$30.2 &    17.6 &   $-$51.7 & 0.5 & 17 \\
 33793 & 2006-07-29   &  28.1 &   330.3 &   $-$30.2 &    17.8 &   $-$52.7 & 0.5 & 11 \\
 33794 & 2006-07-30   &  28.1 &   329.7 &   $-$30.7 &    16.9 &   $-$52.3 & 0.5 & 7 \\
 33795 & 2006-07-30   &  28.1 &   329.7 &   $-$29.7 &    18.5 &   $-$52.2 & 0.5 & 9 \\
 33796 & 2006-07-30   &  28.1 &   329.1 &   $-$30.2 &    17.6 &   $-$51.7 & 0.5 & 14 \\
 33797 & 2006-07-30   &  28.1 &   330.3 &   $-$30.2 &    17.8 &   $-$52.7 & 0.5 & 20 \\
 33798 & 2006-07-30   &  28.1 &   329.7 &   $-$30.7 &    16.9 &   $-$52.3 & 0.5 & 27 \\
 33799 & 2006-07-30   &  28.1 &   329.7 &   $-$29.7 &    18.5 &   $-$52.2 & 0.5 & 33 \\
 33800 & 2006-07-30   &  28.1 &   329.1 &   $-$30.2 &    17.6 &   $-$51.7 & 0.5 & 40 \\
 33801 & 2006-07-30   &  28.1 &   330.3 &   $-$30.2 &    17.8 &   $-$52.7 & 0.5 & 46 \\

\\ \multicolumn{7}{l}{\textit{\underline{\pks (steady) observations (\dataRunsPksSteady~runs)}}} \\[3pt]
 47802 & 2008-08-27   &  28.1 &   330.3 &   $-$30.2 &    17.8 &   $-$52.7 & 0.5 & 36 \\
 47803 & 2008-08-27   &  28.1 &   329.1 &   $-$30.2 &    17.6 &   $-$51.7 & 0.5 & 30 \\
 47804 & 2008-08-27   &  28.1 &   329.7 &   $-$29.7 &    18.5 &   $-$52.2 & 0.5 & 23 \\
 47827 & 2008-08-28   &  28.1 &   330.3 &   $-$30.2 &    17.8 &   $-$52.7 & 0.5 & 35 \\
 47828 & 2008-08-28   &  28.1 &   329.1 &   $-$30.2 &    17.6 &   $-$51.7 & 0.5 & 29 \\
 47829 & 2008-08-28   &  28.1 &   329.7 &   $-$30.7 &    16.9 &   $-$52.3 & 0.5 & 22 \\

\\ \multicolumn{7}{l}{\textit{\underline{\msh observations (\dataRunsMsh~runs)}}} \\[3pt]
 20136 & 2004-03-26   &  28.0 &   228.6 &   $-$58.8 &   320.6 &    $-$0.9 & 0.4 & 38 \\
 20137 & 2004-03-26   &  15.0 &   228.6 &   $-$59.8 &   320.0 &    $-$1.7 & 0.6 & 40 \\
 20151 & 2004-03-27   &  28.1 &   228.6 &   $-$58.8 &   320.6 &    $-$0.9 & 0.4 & 37 \\
 20282 & 2004-04-14   &  28.1 &   228.6 &   $-$58.8 &   320.6 &    $-$0.9 & 0.4 & 37 \\
 20283 & 2004-04-15   &  28.1 &   228.6 &   $-$59.8 &   320.0 &    $-$1.7 & 0.6 & 36 \\
 20301 & 2004-04-15   &  28.1 &   228.6 &   $-$58.8 &   320.6 &    $-$0.9 & 0.4 & 36 \\
 20302 & 2004-04-16   &  28.0 &   228.6 &   $-$59.8 &   320.0 &    $-$1.7 & 0.6 & 36 \\
 20303 & 2004-04-16   &  28.0 &   228.6 &   $-$58.8 &   320.6 &    $-$0.9 & 0.4 & 36 \\
 20322 & 2004-04-16   &  28.0 &   228.6 &   $-$59.8 &   320.0 &    $-$1.7 & 0.6 & 36 \\
 20323 & 2004-04-17   &  28.0 &   228.6 &   $-$58.8 &   320.6 &    $-$0.9 & 0.4 & 36 \\
 20324 & 2004-04-17   &  28.1 &   228.6 &   $-$59.8 &   320.0 &    $-$1.7 & 0.6 & 36 \\
 20325 & 2004-04-17   &  28.0 &   228.6 &   $-$58.8 &   320.6 &    $-$0.9 & 0.4 & 36 \\
 20343 & 2004-04-17   &  28.0 &   228.6 &   $-$58.8 &   320.6 &    $-$0.9 & 0.4 & 37 \\
 20344 & 2004-04-17   &  28.0 &   228.6 &   $-$59.8 &   320.0 &    $-$1.7 & 0.6 & 36 \\
 20345 & 2004-04-18   &  28.1 &   228.6 &   $-$58.8 &   320.6 &    $-$0.9 & 0.4 & 36 \\
 20346 & 2004-04-18   &  28.1 &   228.6 &   $-$59.8 &   320.0 &    $-$1.7 & 0.6 & 36 \\
 20365 & 2004-04-18   &  28.1 &   228.6 &   $-$59.8 &   320.0 &    $-$1.7 & 0.6 & 36 \\
 20366 & 2004-04-18   &  28.1 &   228.6 &   $-$58.8 &   320.6 &    $-$0.9 & 0.4 & 36 \\
 20367 & 2004-04-19   &  28.0 &   228.6 &   $-$59.8 &   320.0 &    $-$1.7 & 0.6 & 36 \\
 20368 & 2004-04-19   &  28.1 &   228.6 &   $-$58.8 &   320.6 &    $-$0.9 & 0.4 & 37 \\

\\ \multicolumn{7}{l}{\textit{\underline{\rxj observations (\dataRunsRxj~runs)}}} \\[3pt]
 20326 & 2004-04-17   &  28.1 &   259.3 &   $-$39.8 &   347.7 &    $-$1.0 & 0.7 & 18 \\
 20327 & 2004-04-17   &  28.1 &   257.5 &   $-$39.8 &   346.9 &     0.1 & 0.7 & 16 \\
 20349 & 2004-04-18   &  28.0 &   259.3 &   $-$39.8 &   347.7 &    $-$1.0 & 0.7 & 16 \\
 20350 & 2004-04-18   &  28.0 &   257.5 &   $-$39.8 &   346.9 &     0.1 & 0.7 & 18 \\
 20396 & 2004-04-20   &  28.1 &   258.4 &   $-$39.1 &   347.9 &    $-$0.1 & 0.7 & 16 \\
 20397 & 2004-04-20   &  28.0 &   258.4 &   $-$40.5 &   346.8 &    $-$0.9 & 0.7 & 19 \\
 20421 & 2004-04-21   &  28.1 &   258.4 &   $-$40.5 &   346.8 &    $-$0.9 & 0.7 & 16 \\
 20422 & 2004-04-21   &  28.1 &   258.4 &   $-$39.1 &   347.9 &    $-$0.1 & 0.7 & 19 \\
 20517 & 2004-04-24   &  28.1 &   257.5 &   $-$39.8 &   346.9 &     0.1 & 0.7 & 18 \\
 20518 & 2004-04-24   &  28.0 &   258.4 &   $-$39.1 &   347.9 &    $-$0.1 & 0.7 & 16 \\
 20519 & 2004-04-24   &  28.0 &   258.4 &   $-$40.5 &   346.8 &    $-$0.9 & 0.7 & 16 \\
 20521 & 2004-04-24   &  28.1 &   259.3 &   $-$39.8 &   347.7 &    $-$1.0 & 0.7 & 23 \\
 20898 & 2004-05-21   &  28.1 &   256.9 &   $-$40.5 &   346.1 &     0.0 & 1.3 & 26 \\
 20899 & 2004-05-21   &  28.0 &   257.5 &   $-$39.9 &   346.8 &     0.0 & 0.7 & 21 \\
 20900 & 2004-05-21   &  28.1 &   258.0 &   $-$39.3 &   347.5 &     0.0 & 0.5 & 18 \\

\\ \multicolumn{7}{l}{\textit{\underline{Off data observations (\dataRunsOff~runs)}}} \\[3pt]
 20275 & 2004-04-14   &  28.1 &   187.3 &     2.6 &   289.7 &    64.8 & -- & 36 \\
 20339 & 2004-04-17   &  28.0 &   201.4 &   $-$42.3 &   309.6 &    20.1 & -- & 24 \\
 20561 & 2004-04-26   &  28.1 &   232.4 &   $-$38.2 &   334.1 &    14.9 & -- & 27 \\
 20734 & 2004-05-13   &  28.0 &   225.0 &   $-$41.9 &   327.1 &    14.8 & -- & 37 \\
 20915 & 2004-05-22   &  28.0 &   186.8 &     2.1 &   288.8 &    64.3 & -- & 29 \\
 21613 & 2004-07-15   &  28.0 &   349.8 &   $-$42.6 &   347.4 &   $-$65.7 & -- & 21 \\
 21753 & 2004-07-21   &  28.0 &   349.8 &   $-$42.6 &   347.4 &   $-$65.7 & -- & 19 \\
 21807 & 2004-07-24   &  28.0 &   343.7 &   $-$27.9 &    24.4 &   $-$64.2 & -- & 9 \\
 21824 & 2004-07-25   &  28.0 &   356.4 &   $-$14.3 &    69.5 &   $-$70.0 & -- & 9 \\
 21851 & 2004-07-28   &  28.0 &    27.1 &    13.5 &   143.5 &   $-$47.0 & -- & 42 \\
 22022 & 2004-08-13   &  20.0 &   255.7 &   $-$48.3 &   339.3 &    $-$4.0 & -- & 43 \\
 22593 & 2004-09-20   &  28.1 &    80.0 &   $-$45.3 &   251.0 &   $-$34.6 & -- & 39 \\
 22997 & 2004-10-11   &  28.1 &    40.2 &    $-$0.0 &   171.5 &   $-$52.3 & -- & 24 \\
 23040 & 2004-10-14   &  28.1 &    68.6 &   $-$47.3 &   253.4 &   $-$42.5 & -- & 24 \\
 23077 & 2004-10-15   &  28.1 &    40.2 &    $-$0.0 &   171.5 &   $-$52.3 & -- & 29 \\
 23143 & 2004-10-21   &  28.1 &    88.3 &   $-$32.3 &   237.7 &   $-$25.7 & -- & 23 \\
 23246 & 2004-11-07   &  28.1 &    67.8 &     5.4 &   190.1 &   $-$27.8 & -- & 30 \\
 23573 & 2004-12-07   &  28.1 &    88.9 &   $-$38.6 &   244.7 &   $-$26.9 & -- & 27 \\
 23635 & 2004-12-13   &  28.1 &    68.3 &     5.9 &   189.9 &   $-$27.1 & -- & 32 \\
 23651 & 2004-12-14   &  28.1 &    68.3 &     4.9 &   190.8 &   $-$27.7 & -- & 37 \\
 23736 & 2005-01-03   &  28.1 &    83.9 &   $-$69.8 &   280.3 &   $-$31.9 & -- & 49 \\
 25345 & 2005-05-04   &  28.0 &   187.3 &     2.6 &   289.7 &    64.8 & -- & 34 \\
 25443 & 2005-05-08   &  27.9 &   187.3 &     2.6 &   289.7 &    64.8 & -- & 26 \\
 25511 & 2005-05-11   &  27.9 &   187.8 &     2.1 &   291.1 &    64.5 & -- & 34 \\
 26077 & 2005-06-04   &  28.2 &   245.0 &   $-$14.9 &   359.7 &    24.2 & -- & 41 \\
 26791 & 2005-06-27   &  20.7 &   233.7 &    24.2 &    37.8 &    53.2 & -- & 52 \\
 26827 & 2005-06-29   &  28.2 &   234.5 &    23.5 &    36.9 &    52.4 & -- & 46 \\
 26850 & 2005-06-30   &  28.2 &   263.4 &   $-$21.5 &     4.9 &     6.2 & -- & 2 \\
 26964 & 2005-07-04   &  28.2 &   262.7 &   $-$20.8 &     5.1 &     7.2 & -- & 12 \\
 27044 & 2005-07-07   &  28.2 &   262.7 &   $-$20.8 &     5.1 &     7.2 & -- & 11 \\
 27121 & 2005-07-10   &  19.9 &   262.7 &   $-$22.2 &     3.9 &     6.4 & -- & 2 \\
 27939 & 2005-08-12   &  28.2 &   355.9 &   $-$14.8 &    67.4 &   $-$70.0 & -- & 9 \\
 27987 & 2005-08-13   &  28.2 &    54.6 &   $-$34.8 &   235.5 &   $-$53.6 & -- & 21 \\
 28341 & 2005-08-31   &  28.2 &   310.0 &    $-$1.6 &    44.6 &   $-$24.8 & -- & 23 \\
 28967 & 2005-09-30   &  28.2 &   310.0 &    $-$1.6 &    44.6 &   $-$24.8 & -- & 23 \\
 28981 & 2005-10-01   &  28.2 &    56.7 &     1.1 &   186.4 &   $-$39.3 & -- & 26 \\
 29024 & 2005-10-02   &  28.2 &    12.7 &   $-$25.3 &   117.4 &   $-$88.2 & -- & 4 \\
 29072 & 2005-10-05   &  28.2 &    11.9 &   $-$26.0 &    85.4 &   $-$88.6 & -- & 20 \\
 29118 & 2005-10-06   &  28.2 &    12.7 &   $-$25.3 &   117.4 &   $-$88.2 & -- & 12 \\
 29177 & 2005-10-09   &  28.2 &    63.7 &     1.1 &   191.5 &   $-$33.6 & -- & 26 \\
 29433 & 2005-10-24   &  28.2 &     9.0 &   $-$71.7 &   304.7 &   $-$45.4 & -- & 50 \\
 29487 & 2005-11-02   &  28.2 &    11.9 &   $-$24.6 &   103.7 &   $-$87.3 & -- & 3 \\
 29526 & 2005-11-04   &  28.2 &     7.3 &   $-$73.0 &   305.2 &   $-$44.0 & -- & 52 \\
 29556 & 2005-11-06   &  28.2 &    64.2 &     0.6 &   192.3 &   $-$33.4 & -- & 25 \\
 29683 & 2005-11-20   &  28.2 &     9.0 &   $-$71.7 &   304.7 &   $-$45.4 & -- & 49 \\

\hline

\end{longtable}
\label{tab:obs_list}
\end{center}

\end{appendix}

\end{document}